\documentclass[useAMS,usenatbib]{mn2e}

\usepackage{amsmath}
\usepackage{amssymb}
\usepackage{graphicx}
\usepackage{natbib}
\usepackage{booktabs}
\usepackage{multirow}
\usepackage{rotating}
\usepackage{subfigure}
\usepackage{xcolor}
\usepackage[innerbars,color]{changebar}
\cbcolor{blue}

\usepackage[T1]{fontenc}
\usepackage{aecompl}

\newcommand{\ud}{\mathrm{d}}
\newcommand{\ue}{\mathrm{e}}

\newcommand{\ui}{\mathrm{i}}

\title[Periodic orbits for 3 and 4 co-orbital bodies]{Periodic orbits for 3 and 4 co-orbital bodies}
\author[P. E. Verrier and C. McInnes]{P. E. Verrier$^{1}$\thanks{E-mail:
patricia.verrier@strath.ac.uk (PEV); colin.mcinnes@strath.ac.uk (CM)} and C. 
McInnes$^{1}$\footnotemark[1]\\
$^{1}$ Advanced Space Concepts Laboratory, Department of Mechanical and Aerospace Engineering, University of Strathclyde}
\begin{document}

\date{}

\pagerange{\pageref{firstpage}--\pageref{lastpage}} \pubyear{}

\maketitle

\label{firstpage}


\begin{abstract}
We investigate the natural families of periodic orbits associated with the equilibrium configurations of the the planar restricted $1+n$ body problem for the case $2\leq n \leq 4$ equal mass satellites. Such periodic orbits can be used to model both trojan exoplanetary systems and parking orbits for captured asteroids within the solar system. For $n=2$ there are two families of periodic orbits associated with the equilibria of the system: the well known horseshoe and tadpole orbits. For $n=3$ there are three families that emanate from the equilibrium configurations of the satellites, while for $n=4$ there are six such families as well as numerous additional connecting families. The families of periodic orbits are all of the horseshoe or tadpole type, and several have regions of neutral linear stability. 
\end{abstract}

\begin{keywords}
celestial mechanics -- planets and satellites: dynamical evolution and stability -- minor planets, asteroids: general
\end{keywords}

\section{Introduction}

The motion of small co-orbital satellites can be studied using the $1+n$ body problem. This is a restricted version of the $n+1$ problem of one central mass with $n$ other masses in a ring surrounding it, which was first studied by \citet{Maxwell1890} as a model of the motion of the rings of Saturn.
 In the restricted version, the $n$ orbiting masses $m_i$ interact with each other, but not the central mass $M$ i.e. the $m_i$ are infinitesimally small and $m_i/M$ tends to zero. The relative equilibria (stationary configurations in a frame rotating with the co-orbiting satellites) of both problems have been extensively studied in the context of central configurations: a configuration where the total (Newtonian) acceleration of each body is proportional to the position vector of the body relative the centre of mass of the configuration. 

The simplest version of the restricted $1+n$ body problem is planar with the co-orbiting satellites in circular orbits with the same mean motion and same mass. In this case the equilibrium configurations lie on a circle centred on the primary mass \citep{Cors2004}. For $2\leq n\leq8$ there are several different equilibrium configurations possible (see Sec. \ref{sec:eq}) \citep{Salo1988, Cors2004, RS2004} while for $n\geq9$ no  configurations other than a regular 1+$n$-gon have been found, where the satellites are equally spaced in a circle about the central mass \citep{Cors2004, Salo1988, Casasayas1994}. Numerical evidence has been provided for this (e.g. \citealt{Cors2004, Salo1988}) and analytic proofs for the cases $n=4$ \citep{Albouy2009} and $n \gtrsim \ue^{73}$ \citep{Casasayas1994}, but it remains an open problem \citep{Albouy2012}. In the unrestricted $n+1$ body problem this $n$-gon configuration has been shown to be linearly stable for $n\geq7$ for large mass ratios between the central mass and orbiting satellites \citep{Moeckel1994}, while in the circular restricted $1+n$ body problem numerical investigations have shown that at least one linearly stable configuration exists for $2\leq n\leq 7$ \citep{Salo1988}. The linear stability results in combination with the Lyapunov Centre Theorem indicate that natural families of periodic orbits of the co-orbital satellites exist about these equilibrium configurations for each value of $n$. 

In the simplest case with $n=2$ the problem is superintegrable and there are two well known stationary configurations with the satellites separated by either $60^\circ$ (linearly stable) or $180^\circ$ (linearly unstable). The trajectories about the first are the familiar tadpole orbits ending on a homoclinic connection to the second equilibrium configuration. Beyond this connection the trajectories are horseshoe orbits. Such a co-orbital configuration is observed in the Solar System in the dynamics of the satellites Janus and Epimetheus around Saturn e.g. \citet{Yoder1983, RS2004}. For systems with three or more satellites the dynamics are more complex, for example \citet{Salo1988} look numerically at the dynamics about some of the equilibrium configurations for small values of $n$ and find both periodic motion and chaotic motion occurs. However, little appears to be known in general about the families of periodic orbits associated with these configurations for $n>2$. Since such periodic motion is observed in the Solar System for $n=2$ the problem is worthy of further study. 

In addition to natural satellites, there are two other applications of such orbits in astrophysics: the first is the potential of such orbits to offer solutions for stable `parking' orbits of large numbers of small objects, such as captured asteroids around the Moon (for example the easily retrievable asteroids identified by \citealt{Yarnoz2013}) or the moons of Mars or larger main belt asteroids, for future in-situ resource use. Knowledge of the primary families of periodic orbits associated with the problem, and their stability, form a basis of engineering such solutions. 

The second concerns exoplanetary systems. There has in recent years been significant interest in the idea of co-orbital or trojan planets: that is, planetary systems where two or more planets orbit at the same mean distance from their star. This idea was first explored in detail by \citet{laughlin}, and has subsequently been the subject of numerous other works investigating not only the dynamics, but detectability and formation as well. Almost all concentrate on two co-orbital planets, either both of similar mass or one of Jupiter mass and another of Earth-mass. Although such configurations can remain `stable' for long times (e.g. \citealt{laughlin,Schwarz2009,Schwarz2007}) and numerical models show plausible formation mechanisms (e.g. \citealt{Beauge2007, Giuppone2012}), analysis of photometric and radial velocity data has yet to reveal any such system \citep{Giuppone2012, Madhusudhan2009}.

Much of the work on the dynamics of exoplanetary co-orbital planets considers either two co-orbital eccentric gas giant planets in the general three-body problem or terrestrial planets at the classical triangular Lagrangian points in the eccentric restricted three-body problem. The $1+n$ body problem on the other hand is suitable for non-negligible but low mass ratios and small eccentricities and so is more suited to similar mass terrestrial planets in roughly circular orbits. Such systems have been less discussed in the literature than those with giant planets.

In this work we look at the primary families of periodic orbits emanating from the equilibrium configurations for $2\leq n\leq 4$ equal mass satellites, as well as some secondary connecting families, using the boundary-value problem methods implemented in the numerical continuation software \textsc{Auto} \citep{Doedel1991,Doedel2011}. A summary of the known equilibrium configurations is presented in Section \ref{sec:eq} and details of the periodic orbits associated with them in Section \ref{sec:po}.

\section{Equations of Motion}
\label{sec:eom}

The equations of motion of the circular restricted $1+n$ body problem have been derived by several authors, including \citet{Yoder1983, Salo1988, RS2004}. Here we follow that of \citet{RS2004}. All masses are assumed to be point masses, and the $n$ co-orbital satellites to be in a circular and planar orbit with the same average mean motion $n_0$ and radius $r_0$. In a frame centred on the central mass rotating with the average mean motion the position of a satellite is described by its angular position $\phi_i$ relative to an arbitrary reference direction and the relative displacement from the mean radius $\xi_i = (r_i-r_0)/r_0$. Units can be chosen such that $r_0$, $n_0$ and the mass of the central object $M$ are all unity. The equations of motion are then
\begin{eqnarray}
\dot{\phi}_i &=& -\frac{3}{2} \xi_i \\
\dot{\xi}_i  &=& -2 \sum_{j\neq i} m_j f^\prime(\Delta\phi_{ij}) \label{eq:eom1}
\end{eqnarray}
where $\Delta\phi_{ij} = \phi_i - \phi_j$ and
\begin{eqnarray}
f(\phi)        &=& \cos \phi - \frac{1}{2\left|\sin \frac{\phi}{2}\right|}\\
f^\prime(\phi) &=& \sin \phi \left(\frac{1}{8\left|\sin\frac{\phi}{2}\right|^3}-1  \right). 
\end{eqnarray}
This system also assumes that there are no close approaches between the satellites, otherwise the motion will not remain co-orbital. The problem is also Hamiltonian and the $\xi_i$ are related to the conjugate momenta to the co-ordinates $\phi_i$. There are three integrals of motion for this system:
\begin{eqnarray}
0 &=& \sum_i m_i \xi_i  \label{eq:int1}\\
I &=& \sum_i m_i \phi_i \label{eq:int2}\\
J &=& \sum_i m_i \left(-\frac{3}{4}\xi_i^2 + \sum_{i\neq j} m_j f(\Delta\phi_{ij})  \right).\label{eq:int3}
\end{eqnarray}
The first two correspond to the invariance of the system under rotations and hence the conservation of angular momentum and the last is the energy integral, or Jacobi constant. 

The system \ref{eq:eom1} is a first order dynamical system in $\mathbb{R}^{n}\times\mathbb{T}^n$. However, the integrals of motion given by \ref{eq:int1} and \ref{eq:int2} can be used to reduce the system to $\mathbb{R}^{n-1}\times\mathbb{T}^{n-1}$ by eliminating a pair of the $\theta_i$ and $\xi_i$ coordinates. For convenience a new angular coordinate $\theta_i$ can be defined as 
\begin{equation}
\theta_i = \phi_{i+1} - \phi_0
\end{equation}
for $i=1,2,\dots n$. The angle $\theta_i$ now represents the angular separation of satellites $2,\dots n$ from satellite 1. Given $\theta_1,\dots \theta_{n-1}$ the position of satellite 1 is then constrained through Eq. (\ref{eq:int2}). Similarly Eq. (\ref{eq:int1}) allows the radial displacement $\xi_1$ of satellite 1 to be eliminated from the equations of motion as well with no need for the introduction of any other new co-ordinates. The new equations of motion are for example in the case $n=4$ with equal masses $m_i=m$:
\begin{eqnarray}
\dot\xi_2 &=& 2m\left(- f^\prime(\theta_1) -  f^\prime(\theta_1-\theta_2) -  f^\prime(\theta_1-\theta_3)\right)\\
\dot\xi_3 &=& 2m\left(- f^\prime(\theta_2) +  f^\prime(\theta_1-\theta_2) -  f^\prime (\theta_2-\theta_3)\right)\\
\dot\xi_4 &=& 2m\left(- f^\prime(\theta_3) +  f^\prime(\theta_1-\theta_3) +  f^\prime(\theta_2 - \theta_3)\right)\\
\dot\theta_1 &=& -\frac{3}{2} \left( 2 \xi_2 + \xi_3 +  \xi_4 \right)\\
\dot\theta_2 &=& -\frac{3}{2} \left( \xi_2 + 2 \xi_3 +  \xi_4 \right)\\
\dot\theta_3 &=& -\frac{3}{2} \left( \xi_2 +  \xi_3 + 2  \xi_4 \right).
\end{eqnarray}
Although this is not a canonical transform the numerical continuation does not require the system to be in canonical co-ordinates and this form is convenient to work with. The reduction also does not affect the linear stability: one pair of eigenvalues in the original system will represent the invariance under rotation and it is these that are not present in the reduced system.

\section{Symmetries}

The equations of motion possess a number of symmetries which are relevant to the equilibria and periodic orbits. Firstly, there is a reversing symmetry $(t\to -t, \xi_i \to -\xi_i)$. A second less obvious reversing symmetry is due to the nature of the function $f$, and is a symmetry about a line through the equilibrium configuration. \citet{Albouy2009} look at the $1+4$ body problem and prove that all relative equilibria are symmetric about a symmetry line, and that either two satellites lie on the symmetry line or none do (one of the configuration in fact possess several symmetry lines, and is of both types). \citet{Cors2011} discuss symmetries of the $1+n$ body problem, noting similarly that there are two cases that occur for equal masses: satellites on the symmetry line (one if $n$ is odd or two if $n$ is even) or no satellites on the symmetry line. 

In addition, the case of equal masses has an additional spatial symmetry that is less obvious: $\phi_i \to \phi_i + \frac{2\pi}{n}$ in the coordinates used here. We can write $\sum_i \dot{\xi_i} = 0$ without need for multiplication by $m_i$, implying $\sum_i \xi_i = \mathrm{constant} = 0$ (as the mean radius is assumed to be 1). Thus $\sum_i \dot{\phi}_i = 0$ from Eq. \ref{eq:eom1} and $\sum_i \phi_i = \mathrm{constant} = c$. However, $c$ is a sum of angles and thus $c\in \mathbb{T}$ also. The dynamics depend only on the relative angular separation of each satellite rather than the location relative to the reference direction, so for any give value of $c$ a rotation of the entire configuration by $2\pi/n$ also exists. This additional symmetry is seen by \citet{Cors2011} for equilibria in the $1+3$ body problem when the masses are equal. The consequence of the symmetry are seen in the connections between the families of periodic orbits emanating from the equilibria, and is discussed further in Section \ref{sec:po}. 

\section{Equilibria}
\label{sec:eq}

The requirement that $\dot\theta_i=0$ means that $\xi_i=0$ for all equilibrium configurations so that the satellites are all located on the mean circle with angular separations are given by solving $\dot \xi_i = 0$. \citet{Salo1988} determine the stationary configurations for $n=2$ to 9 for equal mass satellites, using numerical methods for $n$ greater than 3. The configurations for $2\leq n\leq4$ are shown in Table \ref{tab:eqcf} for a fixed value of the angular momentum constant $c$. There are three distinct types of equilibrium configuration, which \citet{Salo1988} label as type I (arranged in a `line'), type II (arranged in an $n$-gon) and type III (arrange in a line with one satellite on the opposite side) with subdivisions a and b for the type I and III cases (as for some higher values of $n$ two variants of these configurations exist). As mentioned, for $n\geq9$ the only configuration known to exist is type II, and it appears to be always linearly stable. In addition to the numerical evidence provided by numerous authors, \citet{Casasayas1994} provides a more general analytic proof that places a bound on $n$ as greater than approximately $\ue^{73}$ for this to be the case.

The three types can be defined as follows: type II is the simple case of equal spacing, and is linearly stable for $7 \leq n\leq 9$ at least. The type II configuration is unaffected by a rotation by $2\pi/n$ up to an interchange of satellites, and has several symmetry lines. Type I has the satellites in an arc with the spacing increasing from the central pairs outwards. It is symmetric with respect to a symmetry line through the configuration. Type III has $n-1$ satellites in a similar arc and the remaining one on the opposite side, and has similar symmetry. For the type I and type III configurations $n-1$ other alignments of the configuration other than that shown in the table also exist. A numerical search for equilibria for $n\leq 20$ was also performed as part of this work and no other configurations were found in agreement with the results of \citep{Salo1988}. 

The problem is Hamiltonian and thus the eigenvalues of the linearized system occur in reciprocal (and complex conjugate) pairs. \citet{Salo1988} provide the linear stability of each configuration, and the spectrum for each point is shown in the table, excluding the pair of eigenvalues which are zero in the full system. Note that the eigenvalues for the type II configuration occur in degenerate pairs.

\begin{table*}
\begin{tabular}{lcrcrcrcr}
\toprule
$n$ & 2 & & 3 & & 4 & \\
    &   & \multicolumn{1}{c}{$\phi_i$} & & \multicolumn{1}{c}{$\phi_i$} & & \multicolumn{1}{c}{$\phi_i$} \\
\midrule
  & \multirow{7}{*}{\includegraphics[width=0.75in]{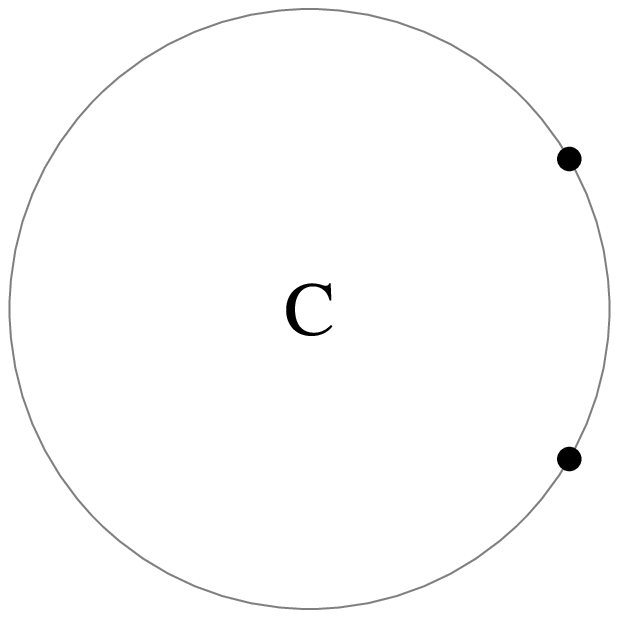}} & & \multirow{7}{*}{\includegraphics[width=0.75in]{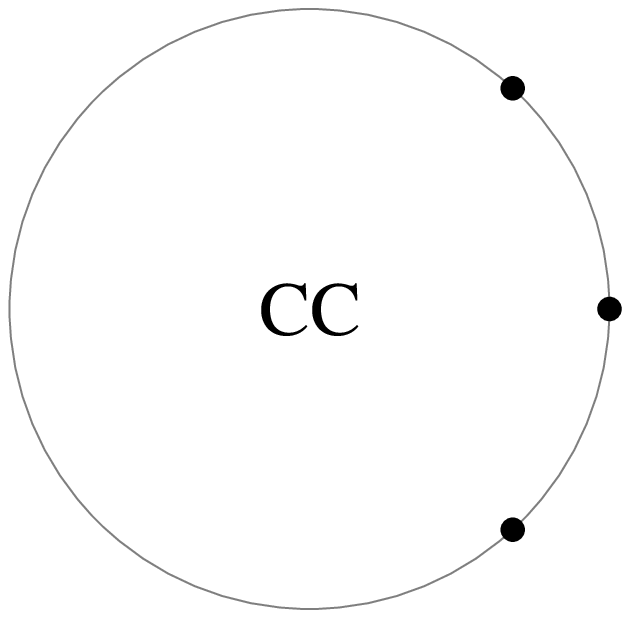}} & & \multirow{7}{*}{\includegraphics[width=0.75in]{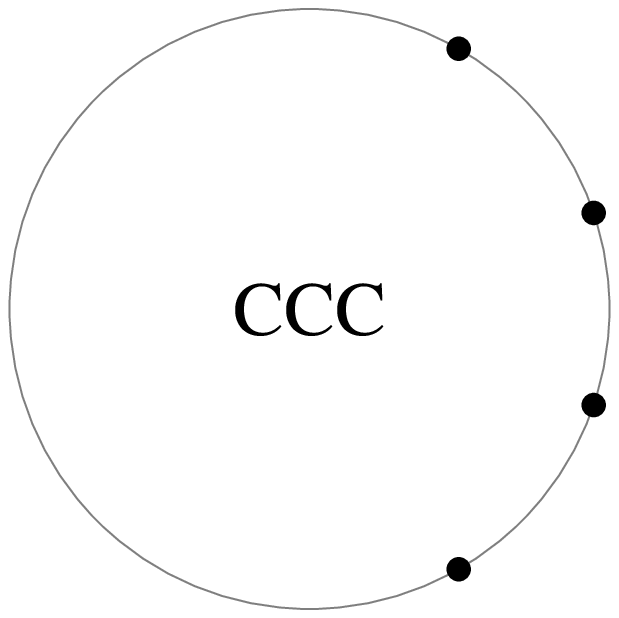}} \\
Ia &  & $30.0^\circ$  & & $0.0^\circ$   & & $18.4^\circ$  \\
   &  & $330.0^\circ$ & & $47.4^\circ$  & & $60.2^\circ$  \\
   &  &               & & $312.6^\circ$ & & $299.8^\circ$ \\
   &  &               & &               & & $341.3^\circ$ \\
   &  &               & &               & &               \\
  \\
  & \multirow{7}{*}{\includegraphics[width=0.75in]{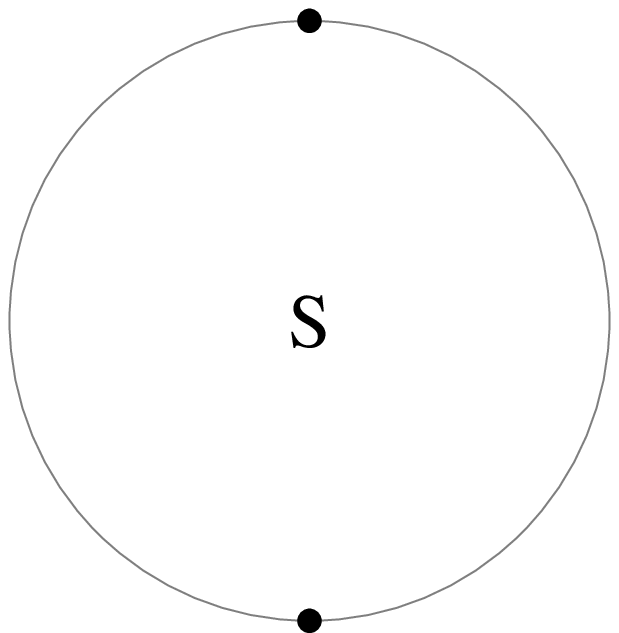}} & & \multirow{7}{*}{\includegraphics[width=0.75in]{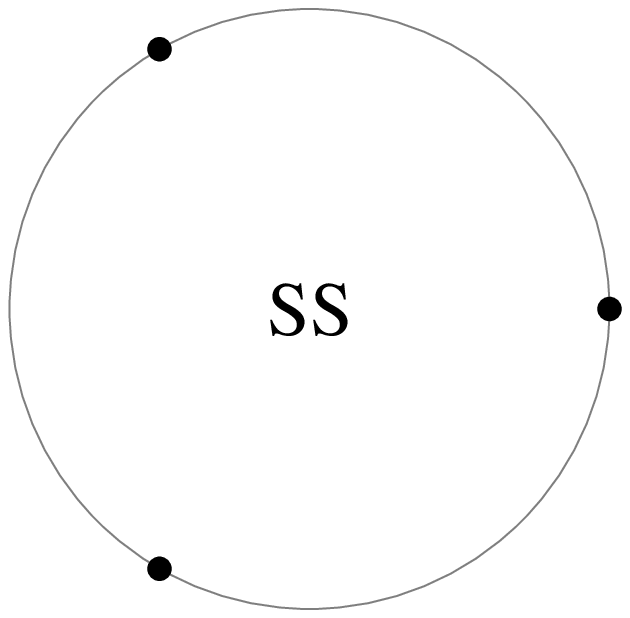}} & & \multirow{7}{*}{\includegraphics[width=0.75in]{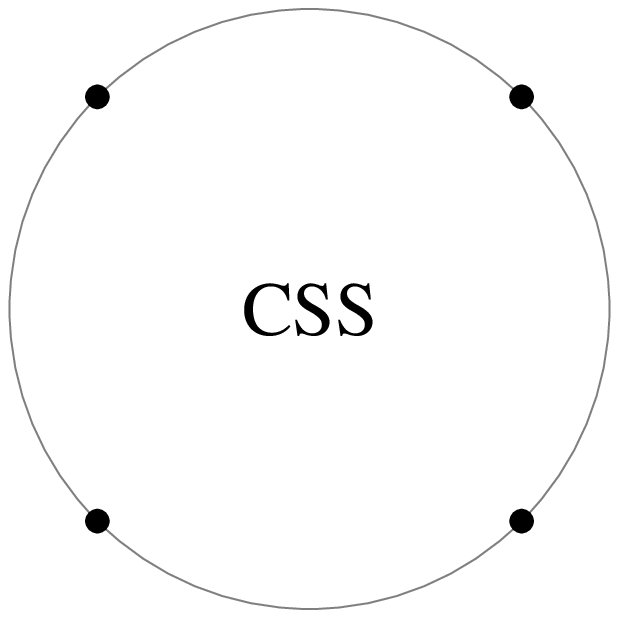}} \\
II &  & $90.0^\circ$  &  & $0.0^\circ$   &  & $45.0^\circ$  \\
   &  & $270.0^\circ$ &  & $120.0^\circ$ &  & $135.0^\circ$ \\
   &  &               &  & $240.0^\circ$ &  & $225.0^\circ$ \\
   &  &               &  &               &  & $315.0^\circ$ \\
   &  &               &  &               &  &               \\
  \\
   &          & & \multirow{7}{*}{\includegraphics[width=0.75in]{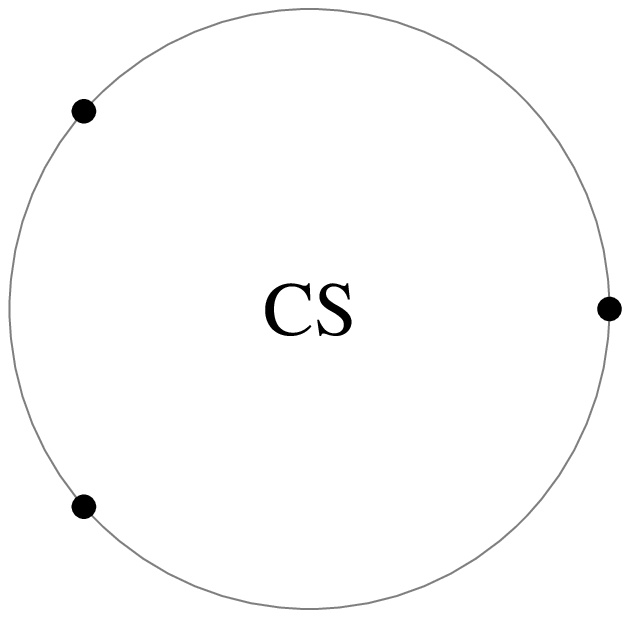}} & & \multirow{7}{*}{\includegraphics[width=0.75in]{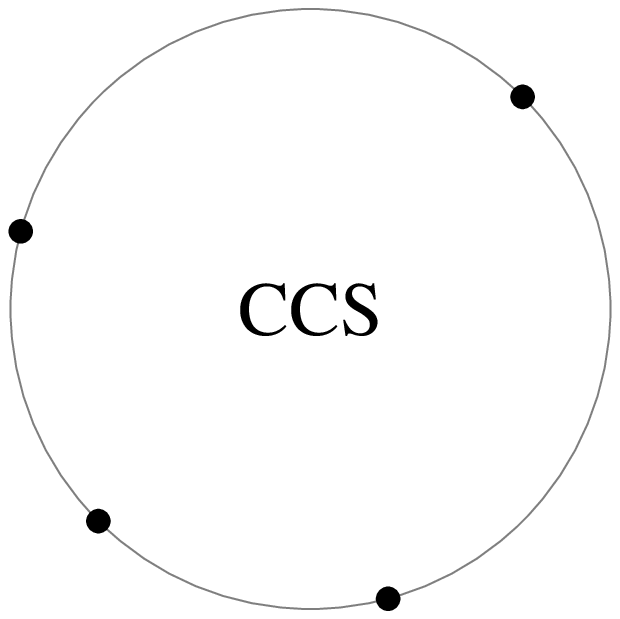}} \\
IIIa  &  &            &  & $0.0^\circ$   &  & $45.0^\circ$  \\
      &  &            &  & $138.8^\circ$ &  & $165.0^\circ$ \\
      &  &            &  & $221.2^\circ$ &  & $225.0^\circ$ \\
      &  &            &  &               &  & $285.0^\circ$ \\
      &  &            &  &               &  &               \\
  \\
\bottomrule
\end{tabular}
\caption{\label{tab:eqcf}Equilibrium configurations and linear stability for $2\leq n\leq 4$, from \citet{Salo1988}. The linear stability is summarised for each configuration: C represents centre and S saddle stability. The angular co-ordinate $\phi_i$ is also given for each configuration, measured anti-clockwise from the positive $x$-direction}
\end{table*}

\section{Numerical Continuation of Periodic orbits}
\label{sec:po}

It is well known that a Hamiltonian system with a first integral has natural families of periodic orbits defined by the value of this integral. The $1+n$ body problem has both angular momentum and energy integrals, however the natural families in the angular momentum integral are trivial: they correspond to the rotation of the periodic orbit about the central mass. The natural families corresponding to the energy integral $J$ are not trivial, and it is these which we consider. Further, the Lyapunov Centre Theorem states that:
\begin{quote}
Let a Hamiltonian system in $d$ degrees of freedom have an equilibrium point $x_0$ with characteristic exponents $\pm \lambda_1, \ldots, \pm\lambda_d$ with $\lambda_i \in \mathbb{C}$. If one of the exponents, say $\lambda_1$, is purely imaginary and the ratio $\lambda_i/\lambda_1$ is not an integer for all $i$ then there is a one-parameter family of periodic orbits emanating from the equilibrium point, and its period tends to $2\pi/|\lambda_1|$ at $x_0$.
\end{quote}
Thus in the reduced system there are one-parameter natural families of periodic orbits emanating from equilibrium configurations that possess pairs of (non-resonant) purely imaginary eigenvalues. From the linear stability results of \citet{Salo1988} (summarized in Table \ref{tab:eqcf}) we can see that there will be one family associated with the Ia configuration for $n=2$, whereas for $n=3$ two families emanate from the Ia configuration and one from the IIIa configuration. For $n=4$ three families will emanate from the type Ia configuration, one from the type II configuration and two from the type IIIa configuraiton. (Note that there may be other families in these systems not associated with the equilibria: for example the well known horseshoe orbits in the $n=2$ case.)

Natural families of periodic orbits can be continued in the boundary-value problem (BVP) numerical continuation software \textsc{Auto} \citep{Doedel1991,Doedel2011}. \textsc{Auto} uses polynomial collocation in association with pseudo-arclength continuation to solve a BVP formulation of the system. Briefly, the problem of finding a periodic orbit $\boldsymbol{x}$ with unknown period $T$ of the first order system $\dot{\boldsymbol{x}}(\tau)=f(\boldsymbol{x}(\tau),p)$ with parameter p can be written as the BVP: 
\begin{eqnarray}
\dot{\boldsymbol{x}}(t) &=& T f(\boldsymbol{x}(t),p) \\
0 &=& \boldsymbol{x}(0) - \boldsymbol{x}(1)\\
0 &=& \int_0^1 \dot{\boldsymbol{x}}^T_{\mathrm{ref}}\boldsymbol{x}(t)\ud t\\
s &=& \int_0^1 (\boldsymbol{x}_\mathrm{ref}(t)-\boldsymbol{x}(t))^T \boldsymbol{x}_\mathrm{tan}(t) \ud t \nonumber\\
{} & & \qquad + (T_\mathrm{ref}-T)T_\mathrm{tan} + (p_\mathrm{ref}-p)p_\mathrm{tan}
\end{eqnarray}
where time has been rescaled so that the periodic orbit has a period of 1. The first equation is the equations of motion, the second the periodicity constraint and the third a phase condition which fixes the phase of the periodic orbit $\boldsymbol{x}$ relative to a reference solution $\boldsymbol{x}_\mathrm{ref}$ (usually the last periodic orbit along the branch). The final equation is the pseudo-arclength constraint, which permits the family to be continued in the parameter $p$. Here $s$ is the step-size and $(\boldsymbol{x}_\mathrm{tan}(\cdot),T_\mathrm{tan}, p_\mathrm{tan})$ the tangent to the reference solution. The use of pseudo-arclength continuation in particular allows families to be continued around folds relative the continuation parameter $p$, ensuring the whole family can be continued. The continuation of a family can be started either by branching directly from an equilibrium point or a branch point in a previously generated family, or by providing a known reference solution. For further details see for example \citet{Doedel2007} or \citet{Munoz2003}.

The natural families of periodic orbits to be considered here are parameterised by the energy level $J$, which does not appear in the equations of motion as formulated. However, a simple solution to this is to introduce an unfolding term into the equations of motion to act as the parameter $p$. 
\citet{Munoz2003} show that a suitable term for systems with a first integral $F$ is $\lambda \nabla F$, where $\lambda$ is an unfolding parameter. For this system the first integral we are interested in is the energy constant $J$. The unfolding term is easily added to the equations of motion, and periodic orbits can be generated directly from an initial equilibrium configuration. If $\boldsymbol{x} = (\theta_{1,\dots,(n-1)},\xi_{2,\dots, n})$ then the system implemented in \textsc{Auto} is
\begin{equation}
\dot{\boldsymbol{x}} = f(\boldsymbol{x}) + \lambda \nabla J
\end{equation}
where $f(\boldsymbol{x})$ are the equations of motion for the reduced system. This BVP formulation and unfolding technique has been applied numerous times in the context of celestial mechanics, for example to the CRTBP by \citet{Doedel2003, Doedel2005,Doedel2007} and \citet{Calleja2012} and to choreographies of the $n$-body problem by \citet{MunozAlmaraz2007}. Given a solution of the reduced system the full system can be calculated by fixing a value of the constant $c$ and using equations \ref{eq:int1} and \ref{eq:int2} to determine the coordinates $(\xi_i, \phi_i)$. Although the system is in $\mathbb{R}^{n-1} \times \mathbb{T}^{n-1}$ in practice the angles do not start to circulate for the periodic orbits presented here and there is no issue with the numerical continuation. 

The dynamics of the system scale with the relative masses of the satellites $m$, and a transform $(t\to\tau/m^{1/2}$, $\xi\to m^{1/2}x)$ removes the dependence completely from Equations \ref{eq:eom1}. In this work we use a very high mass ratio of $\mu=m/M=10^{-2}$ to clearly illustrate the nature of the orbits in the spatial plane only, although realistic mass ratios in the context of the Solar system are much lower, and the dynamics scale accordingly. The periodic orbits are otherwise shown in normalized co-ordinates in all other plots of the motion presented in the following sections.

\subsection{Linear Stability}

The linear stability of a periodic orbit can be determined from its Floquet multipliers. These are the eigenvalues of the monodromy matrix (the state transition matrix evaluated around the orbit) or equivalently the eigenvalues of the linearized Poincar\'e map. The Floquet multipliers of each periodic orbit are calculated by \textsc{Auto}, so information about the linear stability is provided for each continued family. As this is a Hamiltonian system the Floquet multipliers occur in inverse and complex conjugate pairs, with one pair always be equal to unity for a periodic orbit. Hence true linear stability (all multipliers within the unit circle) is not possible in Hamiltonian systems. Instead the linear stability can be described as different orders of instability, depending on how many pairs of multipliers are on the unit circle. For for $n\leq4$ the only possibilities are:
\begin{itemize}
\item order-0 instability (neutral linear stability): all Floquet multipliers on the unit circle (possible for $n\geq2$)
\item order-1 instability: one pair of Floquet multipliers on the real axis (possible for $n\geq3$)
\item order-2 instability: two pairs of Floquet multipliers on the real axis, or in the complex plane but off the unit circle (possible for $n\geq4$).
\end{itemize}

For $n>2$ the floquet multipliers change along a family of periodic orbits, and thus the linear stability of the family can change as well. For $n=3$ or 4 such changes generically occur through either collisions of one pair of multipliers at $+1$ (tangent bifurcation: fold or branch point to another family) or $-1$ (period-doubling bifurcation). For $n=4$ the collision of two pairs of multipliers on the unit circle (Krein collision) or real axis (inverse Krein collision) is also possible.

\subsection{Nomenclature}

As stated in Section \ref{sec:po} the Lyapunov Centre Theorem can be used to determine the existence of natural families emanating from those configurations with (non-resonant) centre eigenvalues. For $n=2,3,4$ there are no resonances between any of the purely imaginary eigenvalues associated with each equilibrium configuration. Thus for a given configuration, each centre frequency (purely imaginary eigenvalues) has a family of periodic orbits associated with it. The linear stability of each configuration is given in Table \ref{tab:eqcf}. So for example, the $n=3$ case has three families that emanate from the equilibria: two associated with the type Ia equilibrium (linear stability centre-centre) and one associated with the type IIIa equilibrium (linear stability centre-saddle). Similarly, the $n=4$ case has six such families: three associated with the centre-centre-centre configuration Ia, one associated with the centre-saddle-saddle configuration II and two associated with the centre-centre-saddle configuration IIIa. Note that the II configuration has no centre frequencies for $n=2$ and $n=3$ and thus there are no families emanating from it. This does not preclude the existence of families not associated with the equilibria, for example horseshoe orbits in the case $n=2$. For a given equilibrium configuration the families associated with each configuration can be labelled as $\alpha$, $\beta$, $\gamma$ and so on, in order of increasing frequency (and thus decreasing period). 

For example, the Ia equilibrium for $n=4$ has three centre frequencies, i.e. it has eigenvalues of the form $\pm \ui \omega_1, \pm \ui \omega_2, \pm \ui \omega_3$, with $\omega_i \in \mathbb{R}$ and $\omega_1 < \omega_2 < \omega_3$. The $\alpha$ family is associated with the $\omega_1$ frequency and has a period that tends to $2\pi/\omega_1$ as the family tends to the equilibrium, the $\beta$ family is associated with the $\omega_2$ frequency has a period  $2\pi/\omega_2$ at the equilibrium and the $\gamma$ family a period $2\pi/\omega_3$ at the equilibrium.

The label for the equilibrium configuration can be added to the end of the designator, as well as a subscript to distinguish the value of $n$. So the family $\mathrm{\alpha Ia_3}$ is the family associated with the lowest centre frequency, emanating from the equilibrium configuration Ia for $n=3$ satellites. A summary of the families considered in this work is shown in Table \ref{tab:pos}.

\begin{table}
\centering
\begin{tabular}{lccccc}
\toprule
n & Ia & Ib & II & IIIa & IIIb \\
\midrule
2 & $\mathrm{\alpha Ia_2}$ & \\
3 & $\mathrm{\alpha Ia_3}$, $\mathrm{\beta Ia_3}$   &  &   & $\mathrm{\alpha IIIa_3}$\\
4 & $\mathrm{\alpha Ia_4}$, $\mathrm{\beta Ia_4}$, $\mathrm{\gamma Ia_4}$ & & $\mathrm{\alpha II_4}$ & $\mathrm{\alpha IIIa_4}$, $\mathrm{\beta IIIa_4}$\\
\bottomrule
\end{tabular}
\caption{\label{tab:pos} Families of periodic orbits associated with the equilibrium configurations from $n=2$ to $n=4$}
\end{table}

\section{Families of periodic orbits}

\subsection{Two co-orbital satellites: $n=2$}

\begin{figure}
\centering
\subfigure[]{
\includegraphics[width=0.98\columnwidth]{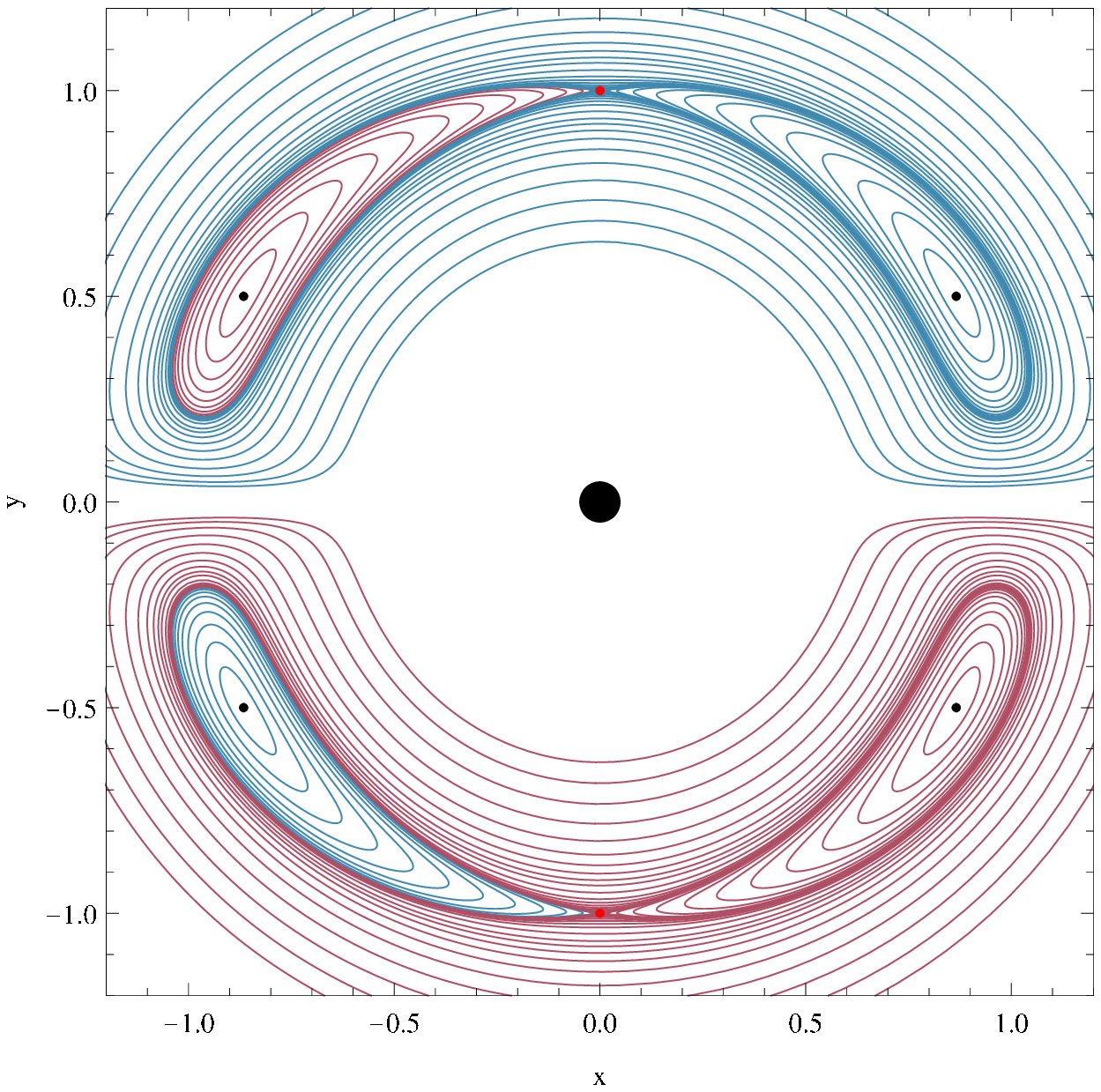}}\\
\subfigure[]{\includegraphics[width=0.98\columnwidth]{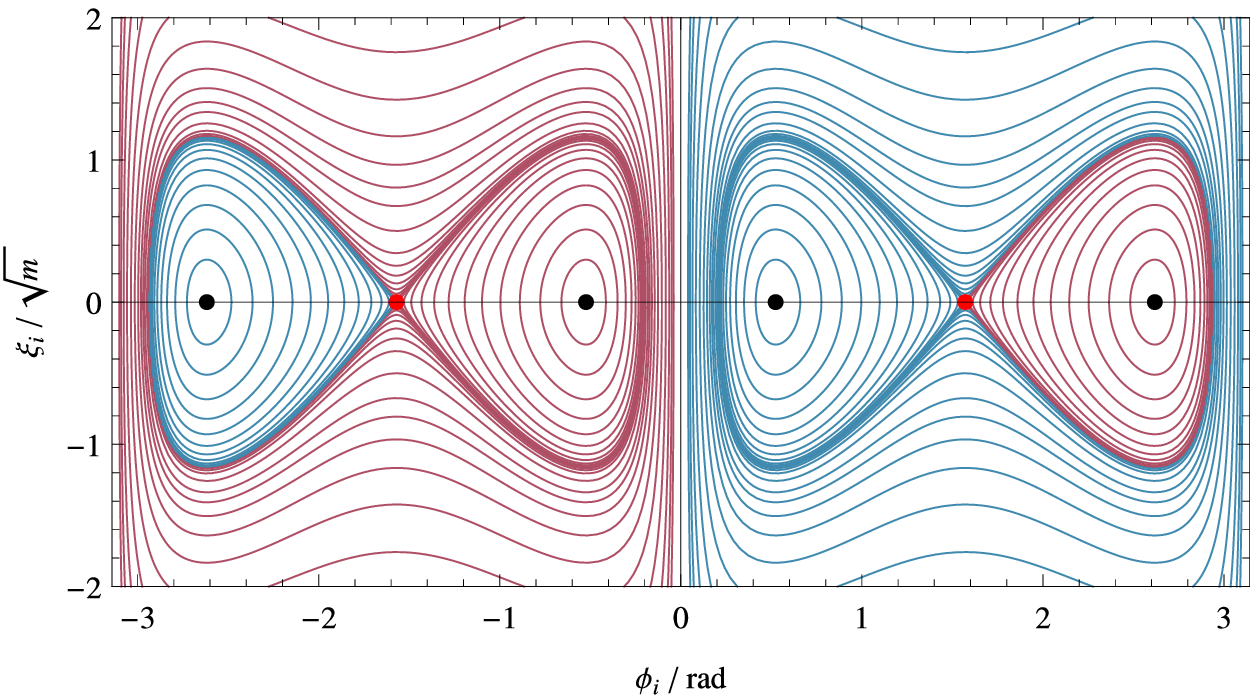}}
\caption{\label{fig:Ia2}The complete dynamics for the $n=2$ case for $c=0$ in the spatial plane and the $\phi_i,\xi_i$ plane. The periodic orbits for one satellite are shown in purple and the other in blue, while the type Ia configuration is shown in black and the type II configuration in red. The horseshoe family continues beyond the amplitudes shown here}
\end{figure}

In the case $n=2$ the system has one degree of freedom and the periodic orbits are well known. They are presented here to give context to the $n=3$ and $n=4$ systems. The families of periodic orbits for this case can be calculated using either the analytic theory of \citet{Yoder1983} (see also \citealt{MD1999}) or by numerical continuation in \textsc{Auto}. The tadpole orbits (the $\alpha \mathrm{Ia}_2$ family) emanate from the type Ia equilibrium and can be generated in  \textsc{Auto} by branching from the equilibrium configuration. The horseshoe orbits on the other hand require an initial orbit to be provided (as they do not emanate from an equilibria), which can be easily calculated from the analytic theory. Both continuation methods were used to generate the families of periodic orbits, and found to agree fully. The two families of periodic orbits, from the \textsc{Auto} continuation, are shown in Figure \ref{fig:Ia2}.

The $\mathrm{Ia_2}$ equilibrium configuration has one family $\mathrm{\alpha Ia_2}$ associated with it, corresponding to tadpole orbits. This family is entirely stable until it terminates on a homoclinic orbit onto type II configuration. Beyond this homoclinic connection is a family of horseshoe orbits. There are two alignments of the $\mathrm{Ia_2}$ equilibrium possible for a given value of $c$: the one shown in the figure and another with the satellites at $\pm 120^\circ$ (assuming the reference direction is along the positive $x$ axis). Thus the full picture of the dynamics includes the rotation about the origin by $180^\circ$ of the family shown in the figure, and horseshoe orbits about the type II configuration, as shown in Figure \ref{fig:Ia2}. This is comparable to the `north' and `south' branches of families of periodic orbits that occur in the CRTBP.

\subsection{Three co-orbital satellites: $n=3$}

\begin{figure*}
\centering
\includegraphics[width=0.3\textwidth]{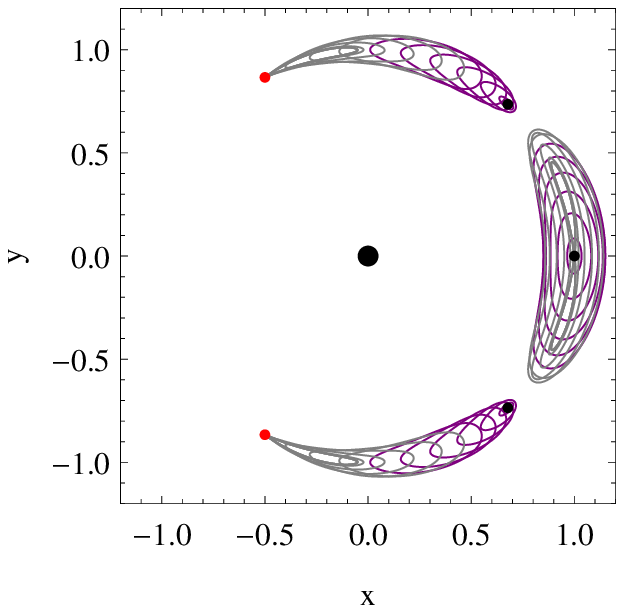}
\includegraphics[width=0.3\textwidth]{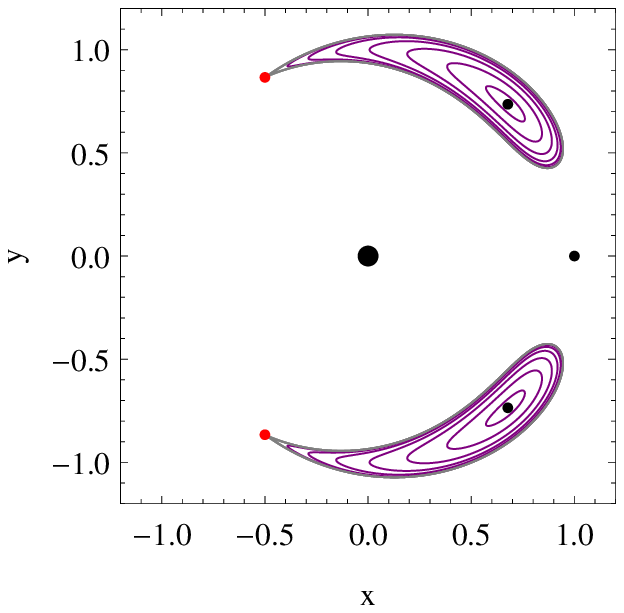}
\includegraphics[width=0.3\textwidth]{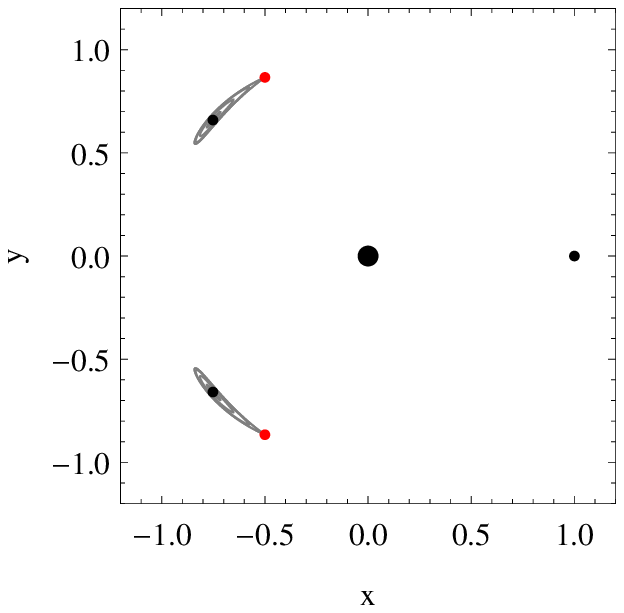}\\
\subfigure[The $\mathrm{\alpha Ia_3}$ family]{\includegraphics[width=0.3\textwidth]{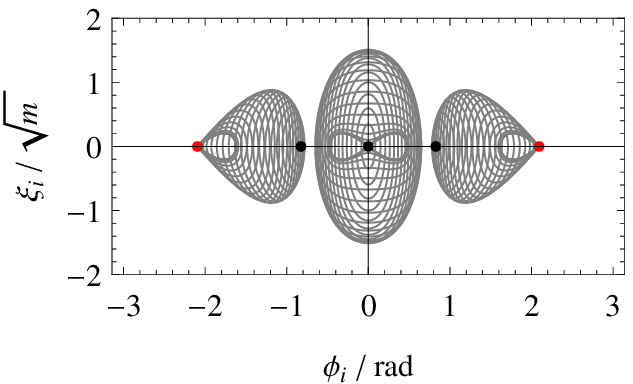}}
\subfigure[The $\mathrm{\beta Ia_3}$ family]{\includegraphics[width=0.3\textwidth]{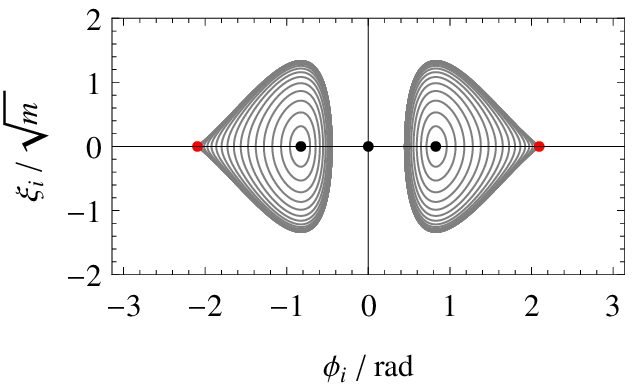}}
\subfigure[The $\mathrm{\alpha IIIa_3}$ family]{\includegraphics[width=0.3\textwidth]{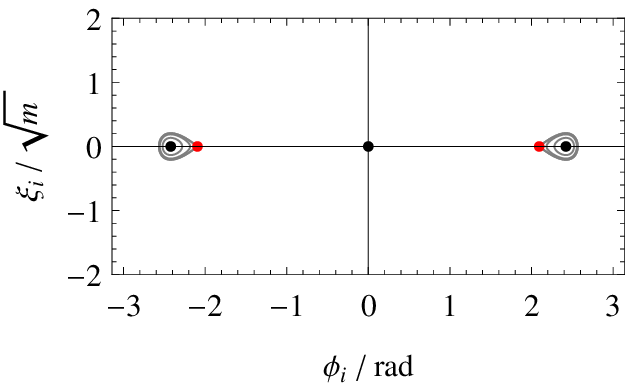}}
\caption{\label{fig:Ia3}The three families of periodic orbits associated with the Ia and IIIa equilibria for $n=3$, in the spatial and $\phi$-$\xi$ planes.
Order-0 instability orbits are shown in purple, order-1 in gray. The relevant location of the Ia or IIIa equilibrium is shown in black and the II equilibrium in red. The configurations all have $c=2\pi$ and the reference direction is taken along the positive $x$ axis}
\end{figure*}

\begin{figure}
\centering
\includegraphics[width=\columnwidth]{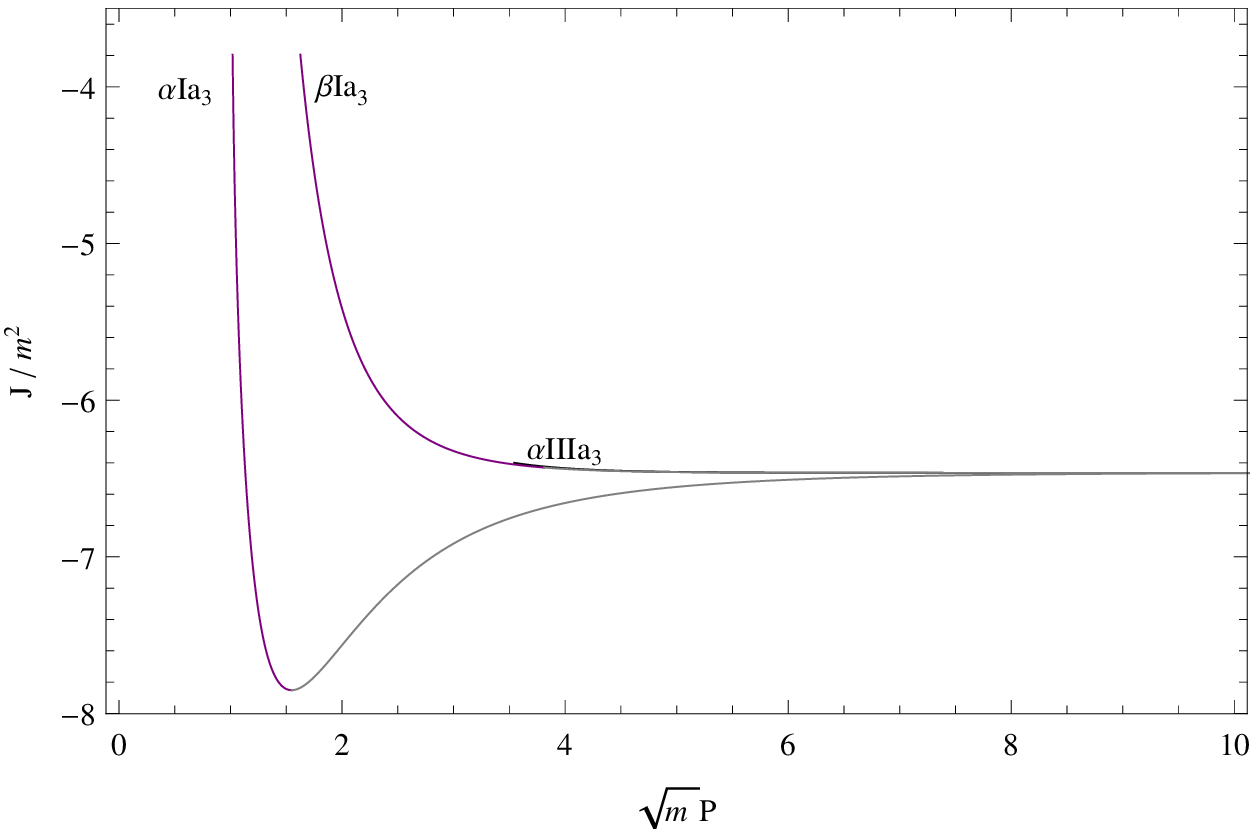}
\caption{\label{fig:n3ep}The three families for $n=3$ plotted in the energy-period plane in units that are scaled to remove the dependence on the mass ratio $\mu=m/M$ and such that $r_0=n_0=M=1$. The order-0 part of the Ia families is shown in purple, and the order-1 part in gray. The IIIa family is shown in black. The families start on the left and all continue out to infinite period}
\end{figure}

For three satellites there are three possible families associated with the equilibria: two emanating from the Ia configuration and one from the IIIa configuration. The two associated with Ia ($\mathrm{\alpha Ia_3}$ and $\mathrm{\beta Ia_3}$) are shown in Fig. \ref{fig:Ia3}. The middle satellite in the $\mathrm{\beta Ia_3}$ family remains stationary ($\xi=\theta=0$) for the entirety of the family. Both families initially have order-0 instability, with $\mathrm{\alpha Ia_3}$ switching to order-1 after a fold in the energy constant and $\mathrm{\beta Ia_3}$ switching to order-1 after a period-doubling bifurcation, and the stable portion is highlighted in Fig. \ref{fig:Ia3}. Both families terminate on a homoclinic orbit to the $\mathrm{II}_3$ equilibrium. Plots of each family in the period-energy plane are shown in Fig. \ref{fig:n3ep}, scaled to eliminate the mass ratio.

The family $\mathrm{\alpha IIIa_3}$ associated with the type IIIa equilibrium is shown in Figure \ref{fig:Ia3} as well. As for $\mathrm{\beta Ia_3}$ one satellite remains stationary. The family has order-1 instability until it ends on a homoclinic orbit to the $\mathrm{II_3}$ equilibrium. 

Similar to the $n=2$ case, the two equilibrium configurations Ia and IIIa exist in three orientations for a given value of the constant $c$ as they are not symmetric under rotation by $2\pi/3$. Thus the families associated with them have two additional symmetric branches. This illustrates that the homoclinic orbit that family $\mathrm{\alpha Ia_3}$ is also a terminating orbit of another branch of the $\mathrm{\alpha IIIa_3}$ family: in the spatial plane these tadpole orbits are interior to the homoclinic orbit.

All three families of periodic orbits are symmetric about the line through the centre of the configuration, as would be expected from the symmetric nature of their associated equilibrium configuration. They also possess the reversing symmetry $\xi\to-\xi$, $t\to-t$.

\subsection{Four co-orbital satellites: n=4}

For four satellites there are now three, two and one family associated with the type Ia, II and IIIa equilibrium configurations respectively. In addition, as before the type Ia and IIIa configurations can be rotated by $\pi/2$ for a given value of $c$. Thus there are nine possible equilibrium configurations for any given value of $c$, as the Ia and IIIa equilibria can be rotated three times by $\pi/2$. As for the previous systems with $n=2$ and $n=3$ this affects the connections between families of periodic orbits.

\subsubsection{The Ia families}

There are three families associated with the Ia equilibrium for this value of $n$. The first family $\mathrm{\alpha Ia_4}$ connects to the other alignments of the $\mathrm{Ia_4}$ configurations via a branch point that is also a connection to the higher symmetry family $\mathrm{\alpha II_4}$ associated with the $\mathrm{II_4}$ configuration. Further, this is a degenerate branch point, as the family $\mathrm{\alpha IIIa_4}$ also connects to $\mathrm{\alpha II_4}$ at the same point. One branch of the $\mathrm{Ia_4}$ family is shown in Fig. \ref{fig:n4families}, with the branching orbit plotted in blue. 

The connection of $\mathrm{\alpha Ia_4}$ and $\mathrm{\alpha IIIa_4}$ to the $\mathrm{\alpha II_4}$ family is similar to the symmetry-breaking pitchfork bifurcations that occur in the CRTBP (e.g. the north and south branches of the Halo family from the planar Lyapunov family). The four branches could be considered to be the north, east, south and west branches of the $\mathrm{\alpha Ia_4}$ family. Similarly in the $n=2$ case the $\mathrm{Ia_2}$ family has two branches, while in the $n=3$ case the $\mathrm{Ia_3}$ and $\mathrm{IIIa_3}$ families have three branches associated with the $\phi\to\phi+\frac{2\pi}{n}$ symmetry.

The family $\mathrm{\beta Ia_4}$ is also shown Fig. \ref{fig:n4families}. It ends in a connection to $\mathrm{\alpha Ia_4}$ via a reverse period-doubling bifurcation. An additional family (labelled $\beta1$) branches off $\mathrm{\beta Ia_4}$ and reconnects further along the family. This family branches off in a symmetry-breaking pitchfork bifurcation, the new family losing the reversing symmetry $\xi\to-\xi$.

The family $\mathrm{\gamma Ia_4}$ appears to connect to $\mathrm{\alpha Ia_4}$ twice: once as a reverse-period tripling bifurcation, and finally ending at a reverse period-quadrupling bifurcation, behaviour very similar to the planar families associated with the L4 and L5 points in the CRTBP. It also has a number of other branch points to four other families, that can be labelled $\gamma1$, $\gamma2$, $\gamma3$, $\gamma4$ in the order they branch from the original family.

\begin{figure*}
\centering
\subfigure[\label{fig:bfdiag1}The connections between the primary families for $n=4$]{
\includegraphics[width=\textwidth]{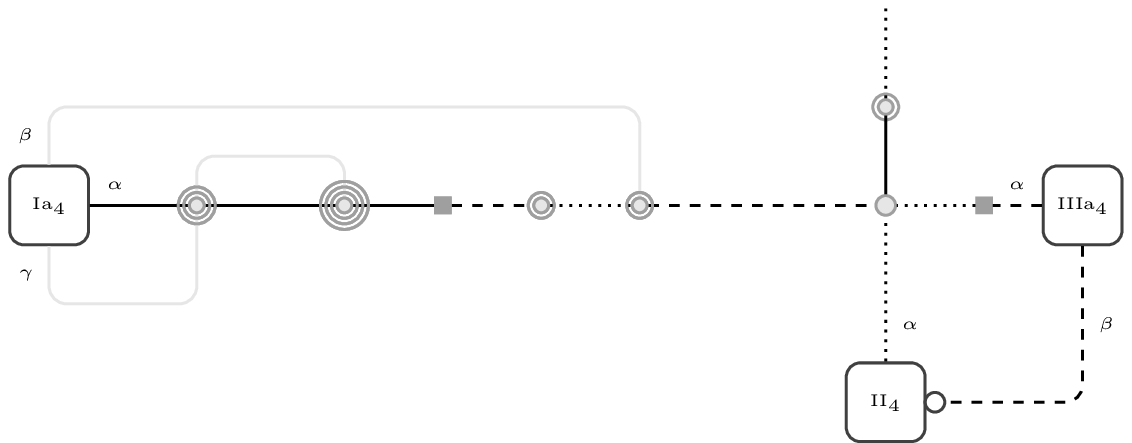}
\vspace{0.2in}
}
\subfigure[\label{fig:bfdiag2}The connections between the $Ia_4$ families]{
\includegraphics[width=\textwidth]{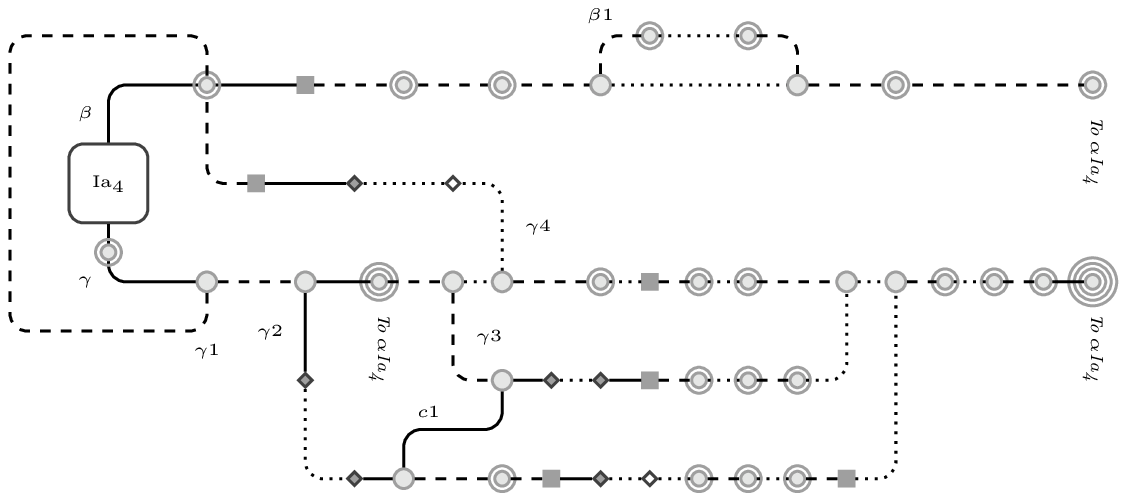}
\vspace{0.2in}}
\subfigure[\label{fig:bfdiag3}Key]{\includegraphics[width=\textwidth]{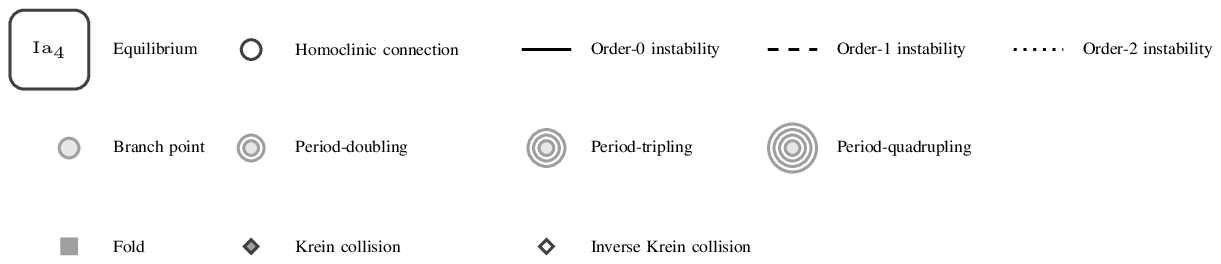}}
\caption{\label{fig:bfdiag} Schematic bifurcation diagrams for the families emanating from the three equilibrium configurations in the $n=4$ system. The top figure shows the connections between the families from the three equilibria. The linear stability and bifurcations along the $\beta\mathrm{Ia}_4$ and $\mathrm{\gamma Ia_4}$ families are not shown in this first diagram. Instead, the middle figure shows these two complicated families and the connections between them in more detail. The key for both diagrams is given in the last figure. For example, the $\alpha\mathrm{IIIa}_4$ family is shown as a horizontal line starting at the equilibrium $\mathrm{IIIa}_4$ represented as an open square on the right-hand side of the top figure. It has one fold (the gray square) before ending at the degenerate branch point (gray circle) with the $\alpha\mathrm{II}_4$ family, represented as the horizontal line starting at the equilibria labelled $\mathrm{II}_4$. The linear stability of the $\alpha\mathrm{IIIa}_4$ family is indicated by the dashing of the line -- initially the family is order-1 (dashed) and changes to order-2 (dotted) after the fold bifurcation. The $\beta\mathrm{IIIa}_4$ family is represented as the other line emanating from the $\mathrm{IIIa}_4$ equilibria, and can be seen to be order-1 instability until it ends on a homoclinic connection to the type II equilibrium (represented as the open circle). The infinite series of branch points and period-doublings (and associated linear stability changes) associated with this connection have been omitted from the diagram}
\end{figure*}

A schematic bifurcation diagram summarizing the connections between the primary families emanating from the Ia equilibrium configuration is shown in Fig. \ref{fig:bfdiag}. The linear stability and the location of various bifurcations are also shown. In the diagram each primary family is represented as a single line emanating from a particular equilibrium configuration, as labelled. (Note that this diagram shows only one branch of each family). The various bifurcations along each family are indicated by various symbols on this line (as defined in the key in Fig. \ref{fig:bfdiag3}). Further, the linear stability of the family is indicated through dashing or dotting of the line representing the family. For example, the $\alpha \mathrm{Ia}_4$ family is represented as the horizontal line labelled $\alpha$, starting at the equilibrium labelled $\mathrm{Ia}_4$ on the left of the diagram (represented as a large square) and ending at the degenerate branch point on the vertical line which represents the $\alpha\mathrm{II}_4$ family. The various bifurcations along the family are shown as indicated, and the linear stability can for example be seen to change from order-0 initially to order-1 after a fold (gray square), and so on.

Fig. \ref{fig:bfdiag1} shows the connections between the primary families from the type Ia, II and IIIa configurations (for a description of the families associated with the II and IIIa configurations see the next two subsections). The $\beta\mathrm{Ia}_4$ and $\gamma\mathrm{Ia}_4$ families are complex and their linear stability and bifurcations are not represented fully on this first schematic diagram. Instead only the points at which they connect to the $\alpha\mathrm{Ia}_4$ family are shown. The full bifurcation diagram for these two families, including the secondary families connecting to the branch points in the Ia families are shown separately in the schematic bifurcation diagram in Fig. \ref{fig:bfdiag2}. Note that whereas these diagrams show all branch points, folds and period-doubling bifurcations, the only period tripling and quadrupling bifurcations that have been detected are those between the three primary families associated with the $\mathrm{Ia}_4$ configuration. The secondary families have been labelled in the order they branch off the primary family. There is one family branching from $\mathrm{\beta Ia_4}$ ($\beta1$) and four from $\mathrm{\gamma Ia_4}$ ($\gamma 1$, $\gamma 2$, $\gamma 3$, $\gamma 4$) as discussed above, as well as one family labelled $c1$ that connects between $\gamma2$ and $\gamma3$. All these secondary families are shown in the Appendix, and it is worth noting that several of them also have some order-0 instability, for example the relatively complex asymmetric family labelled $c1$ is completely neutrally linearly stable.

\subsubsection{The II family}

There is one family associated with the $\mathrm{II_4}$ configuration: a family of symmetric horseshoe-type orbits $\mathrm{\alpha II_4}$. As for its associated equilibrium point, this family of periodic orbits has degenerate Floquet multipliers (the two non-trivial pairs are equal), as evidenced by the double branch point to the $\mathrm{\alpha Ia_4}$ and $\mathrm{\alpha IIIa_4}$ families. The $\mathrm{\alpha II_4}$ family only has one branch (unlike the four branches of the Ia and IIIa families), as it is invariant under rotation by $\pi/2$ up to the interchange of satellites.

Interestingly, some of the horseshoe orbits in the $\mathrm{\alpha II_4}$ family have order-0 instability. This occurs just after the double branch point shown in Fig. \ref{fig:8d}, but is ended shortly later along the family by a period-doubling bifurcation.

\subsubsection{The IIIa families}
\label{sec:IIIa4}

As for the $\mathrm{Ia_4}$ configuration there are four possible alignments of the $\mathrm{IIIa_4}$ configuration, and thus four rotated branches of the two associated families. The first, $\mathrm{\alpha IIIa_4}$, behaves similarly to the $\mathrm{\alpha Ia_4}$ family and connects to the branches from the other alignments of the configuration at the degenerate branch point in the $\mathrm{\alpha II_4}$ family. The second, $\mathrm{\beta Ia_4}$, is a tadpole-like family, and ends on a homoclinic orbit from the $\mathrm{II_4}$ equilibrium. Unlike the $\mathrm{\alpha Ia_4}$ and $\mathrm{\alpha II_4}$ families there are no regions of order-0 stability along either of the $\mathrm{IIIa_4}$ families. Both are initially order-1, with the  $\mathrm{\alpha IIIa_4}$ family transitioning to order-2 after a fold. 

As the $\mathrm{\beta IIIa_4}$ family evolves towards the homoclinic orbit it undergoes an infinite series of pairs of branch points and pairs of period-doubling bifurcations. The pair of Floquet multipliers on the real axis tend to zero and infinity, while the pair on the unit circle circulate with increasing frequency, briefly moving off onto the real axis at each bifurcation. The first two families branching off connect to the $\mathrm{\alpha IIIa_4}$ family through reverse period quadrupling bifurcations. This behaviour is similar to that of the the planar Lyapunov families around the triangular equilibrium points in the CRBTP (see e.g. \citealt{Doedel2007}).

\subsubsection{Summary} 

\begin{figure}
\subfigure[The $\mathrm{\alpha Ia_4}$ family]{\includegraphics[width=0.49\columnwidth]{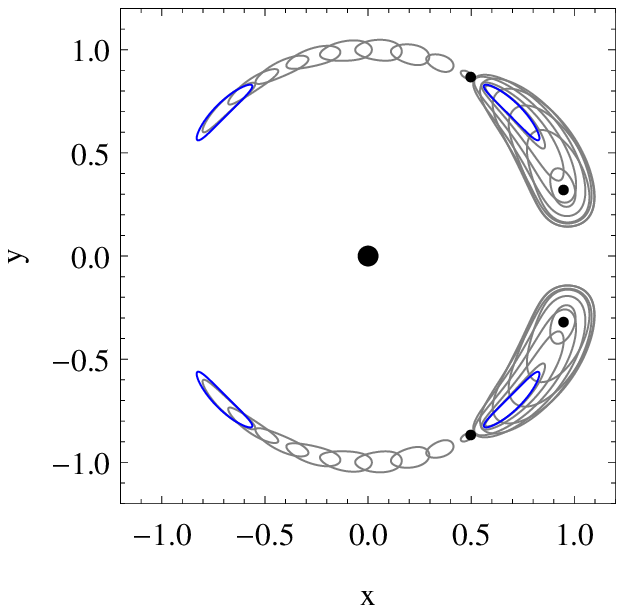}}
\subfigure[The $\mathrm{\beta Ia_4}$ family]{\includegraphics[width=0.49\columnwidth]{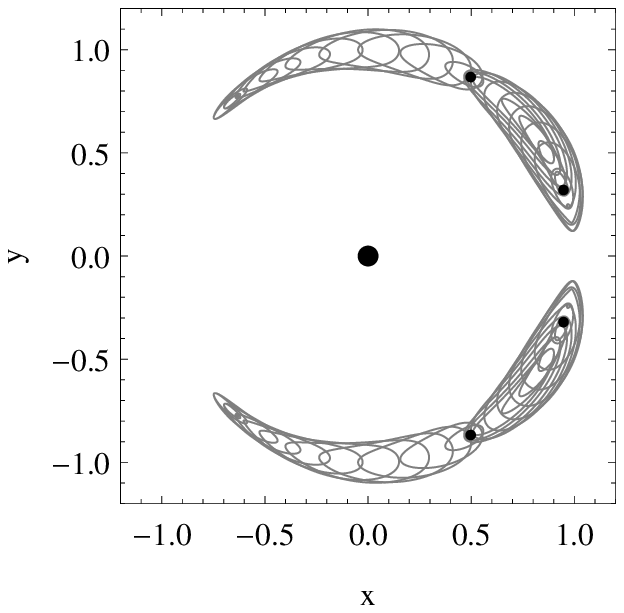}}
\subfigure[The $\mathrm{\gamma Ia_4}$ family]{\includegraphics[width=0.49\columnwidth]{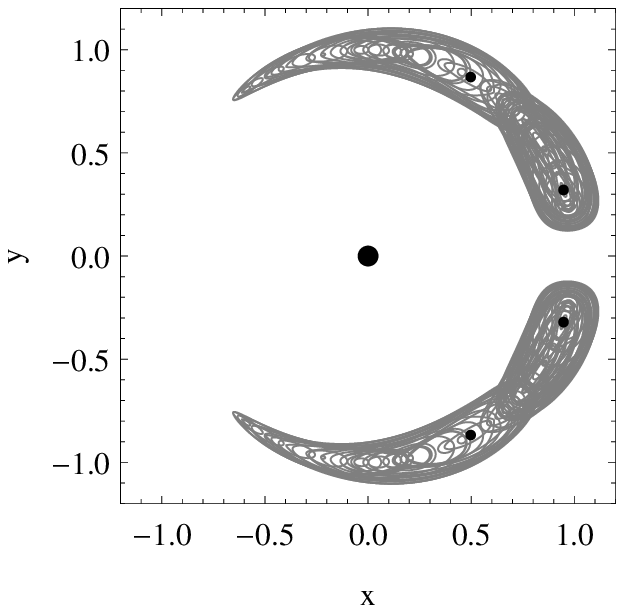}}
\subfigure[\label{fig:8d}The $\mathrm{\alpha II_4}$ family]{\includegraphics[width=0.49\columnwidth]{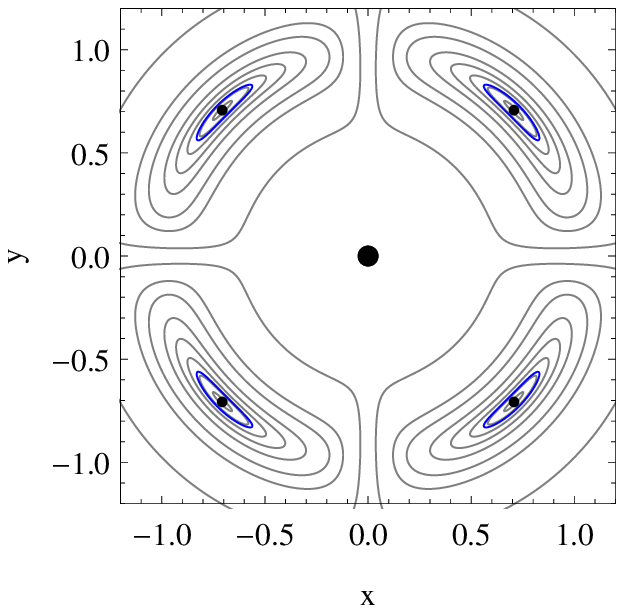}}
\subfigure[The $\mathrm{\alpha IIIa_4}$ family]{\includegraphics[width=0.49\columnwidth]{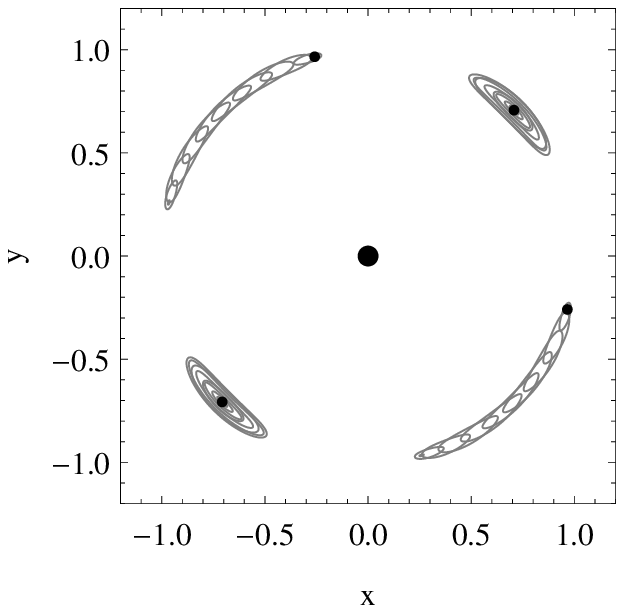}}
\subfigure[The $\mathrm{\beta IIIa_4}$ family (2 of the satellites are stationary)]{\includegraphics[width=0.49\columnwidth]{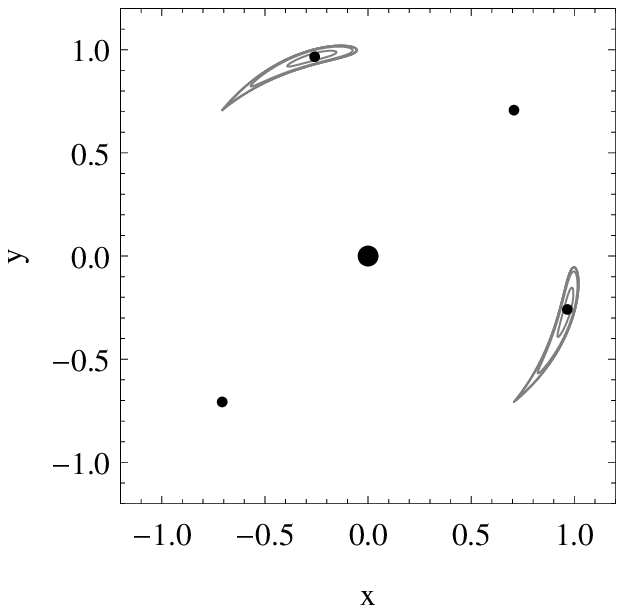}}
\caption{\label{fig:n4families} The families of periodic orbits for $n=4$. The double branch point in the $\mathrm{\alpha II_4}$ family is shown in blue. }
\end{figure}

\begin{figure}
\includegraphics[width=\columnwidth]{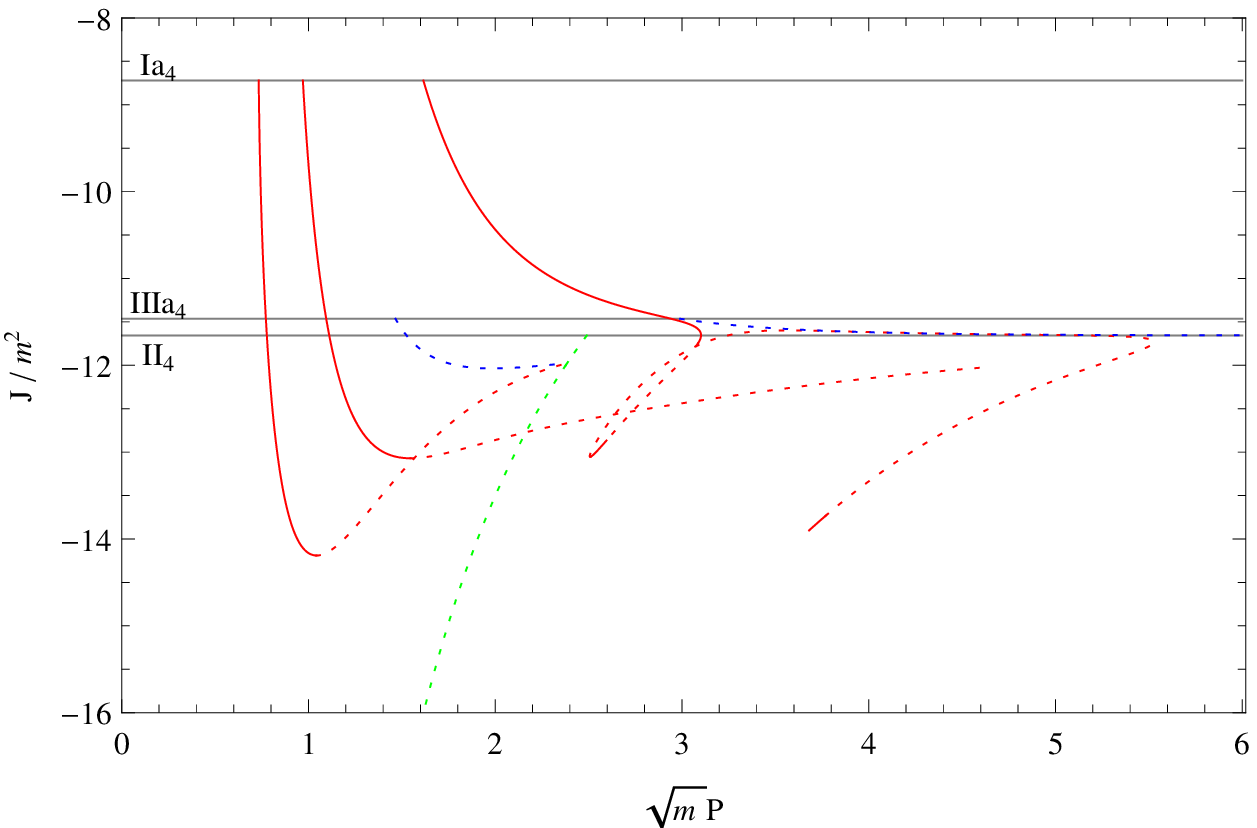}
\caption{\label{fig:stabind} The six primary families emanating from the equilibrium configurations for $n=4$, in the energy-period plane. The Ia familes are shown in red, the II family in green and the IIIa families in blue. The energy level of each equilibrium is also shown, and note that the families are labelled $\alpha$, $\beta$ etc in order of increasing initial period. Solid lines indicated order-0 stability, dotted lines order-1 or order-2 stability}
\end{figure}

All six primary families for $n=4$ are shown in Fig. \ref{fig:n4families} in the spatial plan. In addition, the families are shown in the energy-period plane in Fig. \ref{fig:stabind}. The order-0 portion of each family is highlighted in this plot. Note that the connection of the $\mathrm{\alpha Ia_4}$ and $\mathrm{\alpha IIIa_4}$ families to the $\mathrm{\alpha II_4}$ family can be clearly seen in this plane.

The six primary families are shown plotted in the energy-period plane in Fig. \ref{fig:stabind}. The order-0 portion of each family is again highlighted. The connection of two of the Ia families to the II family is clearly shown in this plane.

The primary families $\mathrm{\alpha Ia_4}$, $\mathrm{\beta Ia_4}$, $\mathrm{\gamma Ia_4}$, $\mathrm{\alpha IIIa_4}$ and $\mathrm{\beta IIIa_4}$ emanating from the Ia and IIIa equilibria are symmetric about the symmetry line of the associated equilibrium configuration. They also possess the reversing symmetry $\xi\to-\xi$, $t\to-t$. The family $\mathrm{\alpha II_4}$ emanating from the type II equilibrium is symmetric about (both) symmetry lines of the equilibrium configuration and is also symmetric under $\phi\to \phi+\frac{2\pi}{n}$. It too has the reversing symmetry $\xi\to-\xi$, $t\to-t$.

The family $\beta1$ branching from $\mathrm{\beta Ia_4}$ is not symmetric about the symmetry line of the equilibrium, but still possess the reversing symmetry $\xi\to-\xi$, $t\to-t$. The families $\gamma1$ and $\gamma3$ branching from $\mathrm{\gamma Ia_4}$ do not have the possess the reversing symmetry $\xi\to-\xi$, $t\to-t$ but are symmetric about the symmetry line of the equilibrium, while the families $\gamma2$ and $\gamma4$ are not symmetric about the symmetry line of the equilibrium but do possess the $\xi$ reversing symmetry. The family $c1$ that connects between branch points in the $\gamma2$ and $\gamma3$ families possesses no symmetries. It should be noted that those families that do not possess one of the reversing symmetries have two symmetric branches, similar to for example the axial families in the CRBTP. Because it lacks both reversing symmetries there are four branches of the $c1$ family, connecting between the two branches of each of the $\gamma2$ and $\gamma3$ families.

\section{Conclusions}
\label{sec:asteroids}

The families associated with the equilibrium configurations of the planar restricted $1+n$-body problem are complex and exhibit a variety of interesting dynamical phenomena, such as homoclinic connections and degenerate symmetry-breaking bifurcations. For $2\leq n\leq4$ the families associated with the type II equilibria resemble classical horseshoe orbits, while some associated with the type I and type III equilibria resemble tadpole orbits. In addition, there are many other periodic orbits, both symmetric and asymmetric exhibiting different linear stability properties (although note that vertical stability of the orbits is not considered here). 

While exactly equal mass satellites and asteroids in the solar system are unlikely to occur, the equilibrium configurations are known to exist in such cases (e.g. \citealt{RS2004}) and as such the periodic orbits presented here provide a starting point for exploring such systems. Preliminary investigation of unequal mass satellites have shown that many families of periodic still exist.

While this work has only considered up to four equal mass satellites the linear stability of the known equilibria for $n>4$ indicate a large number of families of periodic orbits exist in these cases as well. For example, for $n=5$ there are type Ia, II and IIIa equilibria as for the $n=4$ case, and it is found that there are two families emanating from the type II configuration, one of which connects to the type Ia equilibrium point and one of which connects to the type IIIa equilibrium point. The families remain of similar types to those seen in this work: largely tadpole or horseshoe type. 

It should be noted here that the $1+n$-body problem is a restricted problem, and for the periodic orbits to be reproducible in the unrestricted system the mass ratios must be low. While this is not a problem for asteroids around planets or large moons in the Solar System it can become an issue when considering trojan exoplanetary systems. However, direct numerical integration of the full $n$-body equations of motion of an example order-0 periodic orbit from the $c1$ family shows for example four Earth-mass co-orbital planets at 1 au from a solar-mass star are well represented by this model. We also note that such a periodic orbit of four Earth-mass planets would have a period of approximately 250 years, making detection of the co-orbital dynamics difficult. However, the existence of neutrally linearly stable periodic orbits of four trojans still presents an interesting possibility for exoplanetary systems.

The question remains whether such orbits could exist (or be engineered) in the solar system or exoplanetary systems. For co-orbital asteroids, the easily retrievable asteroids suitable for capture into the Earth-Moon system identified by \citet{Yarnoz2013} are at most diameter 65 m, with those that are retrievable in the near-term generally below 4 m diameter, which corresponds to $\mu\approx10^{-14}$ to $\mu\approx10^{-18}$ (assuming an average density of 2 g/cm${^3}$). For these extremely low mass ratios the co-orbital dynamics will be negligible compared to other perturbations. (By comparison, an asteroid with a higher mass ratio of $10^{-6}$ would correspond to an spherical object roughly of diameter 40 km.) This is in some sense a self-solving problem: if the asteroids do not significantly perturb each others orbits, there is no need to consider placing them in a stable parking configuration. It should also be noted that the $1+n$ body problem is not a good approximation of asteroid orbits about the Moon regardless of mass ratio, as the perturbations from the Earth cannot be ignored. The families of periodic orbits found here however provide a starting point for more advanced models of the dynamics. The mechanism by which asteroids could be transferred into such orbits is as well is another topic, but knowledge of the possible orbital configurations will also assist such endeavours.

\section*{Acknowledgments}

This work was funded through the European Research Council, Advanced Investigator Grant 227571 VISIONSPACE. We thank the anonymous referee for helpful comments.

\appendix

\renewcommand{\thefigure}{A\arabic{figure}}

\section*{Appendix}

\begin{figure}
\centering
\subfigure[$\sqrt m P =0.736$]{\includegraphics[width=0.48\columnwidth]{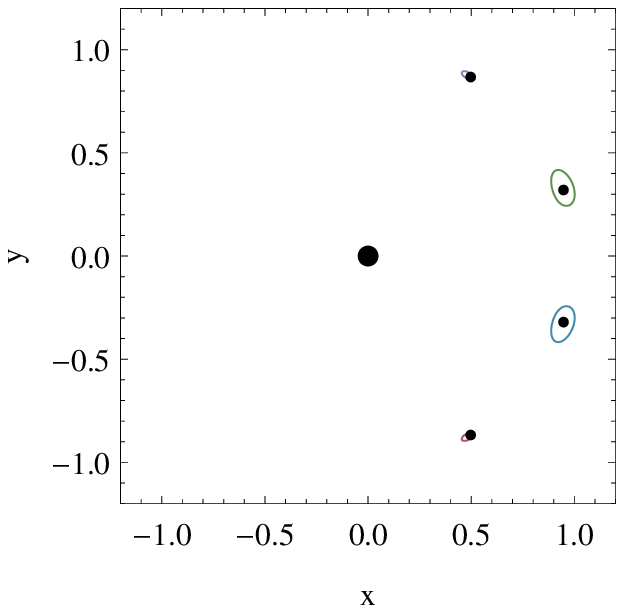}}
\subfigure[$\sqrt m P =0.859$]{\includegraphics[width=0.48\columnwidth]{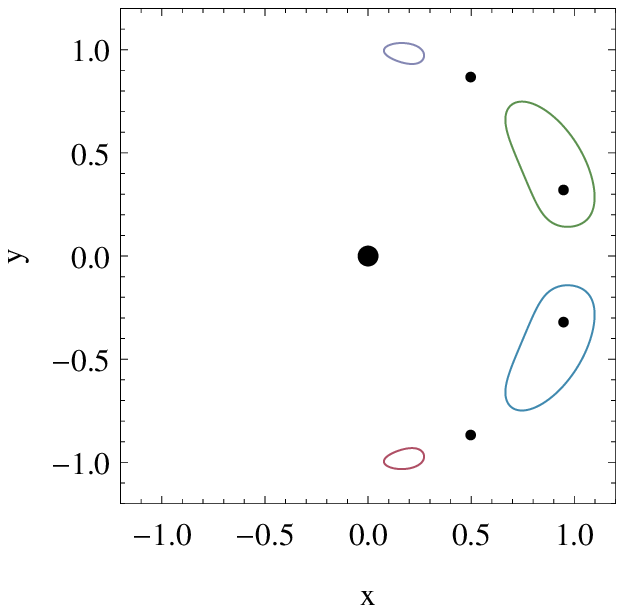}}
\subfigure[$\sqrt m P =1.250$]{\includegraphics[width=0.48\columnwidth]{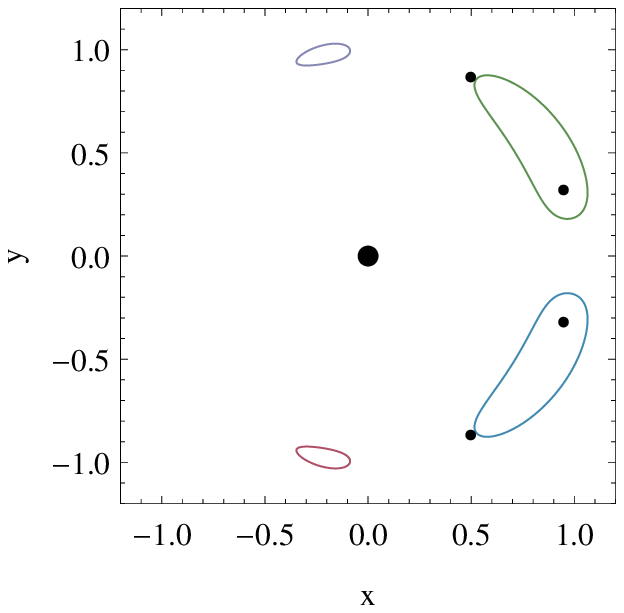}}
\subfigure[$\sqrt m P =2.384$]{\includegraphics[width=0.48\columnwidth]{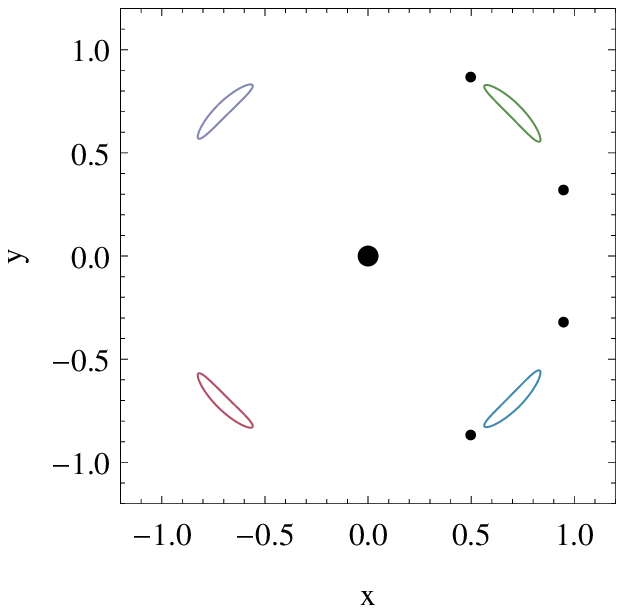}}
\caption{\label{fig:a1}The family $\mathrm{\alpha Ia_4}$}
\end{figure}

\begin{figure}
\centering
\subfigure[$\sqrt m P =3.181$]{\includegraphics[width=0.48\columnwidth]{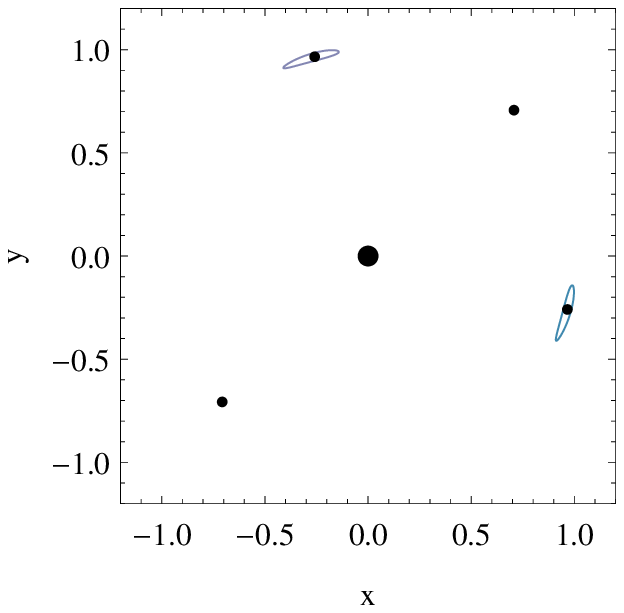}}
\subfigure[$\sqrt m P =6.018$]{\includegraphics[width=0.48\columnwidth]{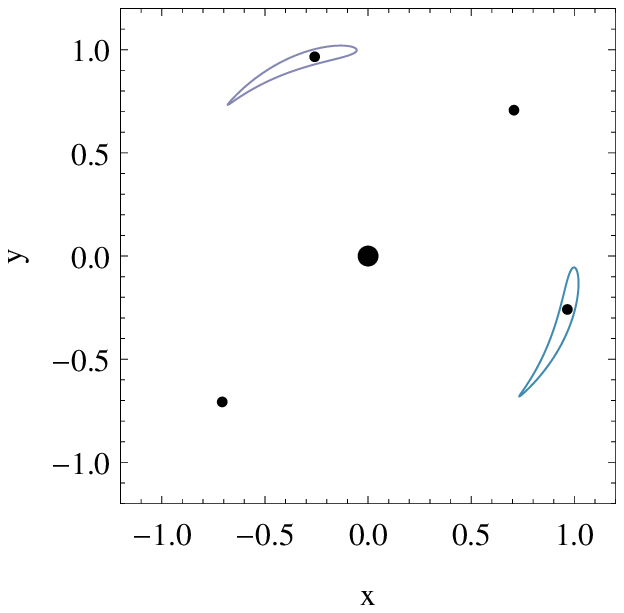}}
\caption{\label{fig:a3}The family $\mathrm{\beta IIIa_4}$}
\end{figure}

\begin{figure}
\centering
\subfigure[$\sqrt m P =3.428$]{\includegraphics[width=0.48\columnwidth]{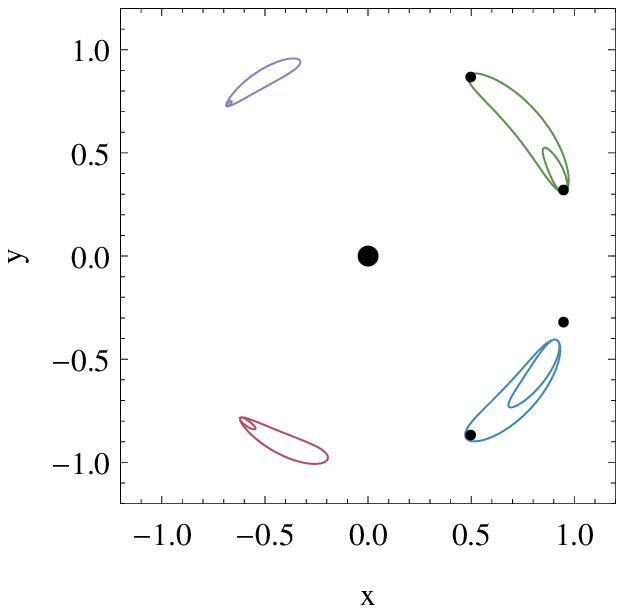}}
\subfigure[$\sqrt m P =4.358$]{\includegraphics[width=0.48\columnwidth]{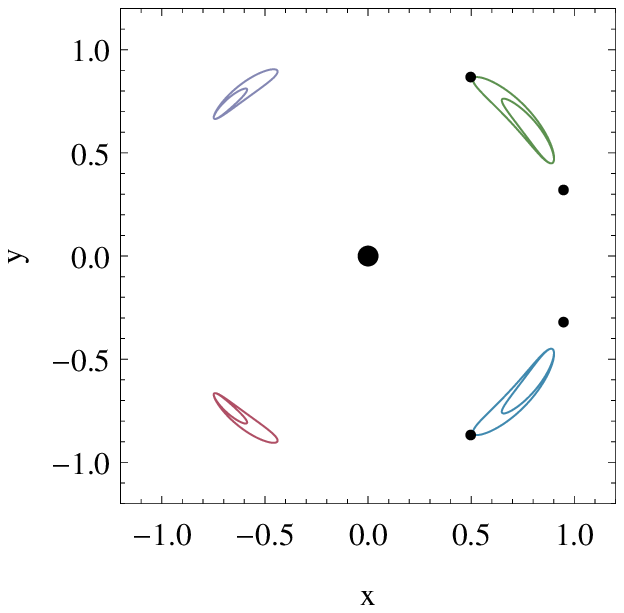}}
\caption{\label{fig:b1}The family $\beta 1$}
\end{figure}

\begin{figure}
\centering
\subfigure[$\sqrt m P =2.710$]{\includegraphics[width=0.48\columnwidth]{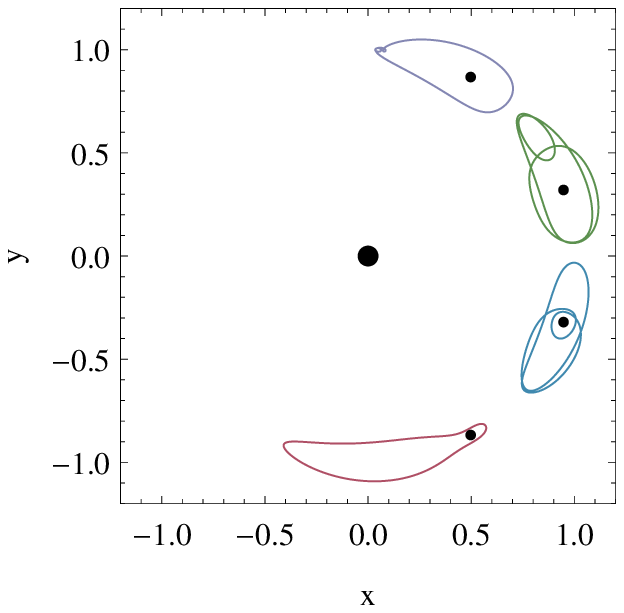}}
\subfigure[$\sqrt m P =2.712$]{\includegraphics[width=0.48\columnwidth]{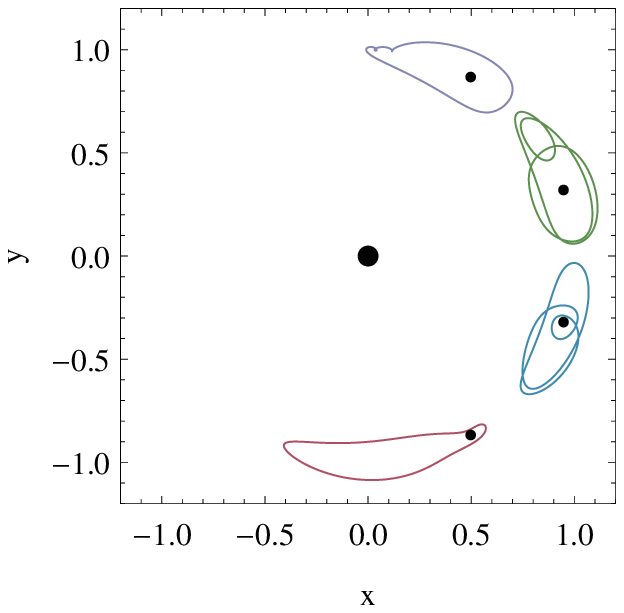}}
\subfigure[$\sqrt m P =2.747$]{\includegraphics[width=0.48\columnwidth]{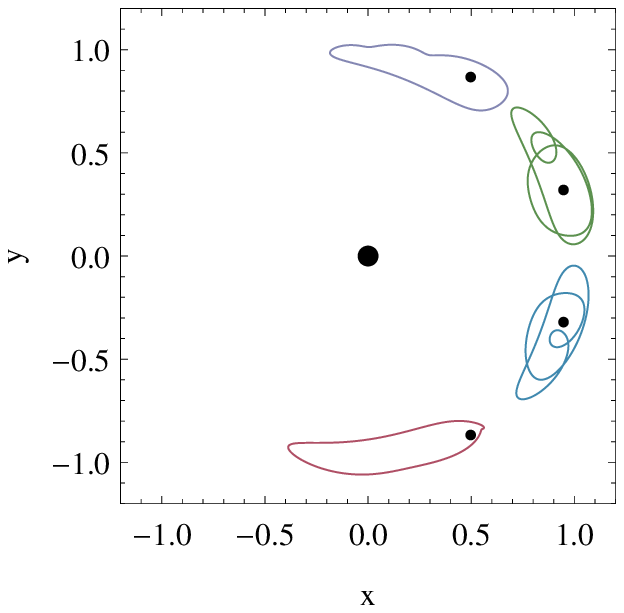}}
\subfigure[$\sqrt m P =2.781$]{\includegraphics[width=0.48\columnwidth]{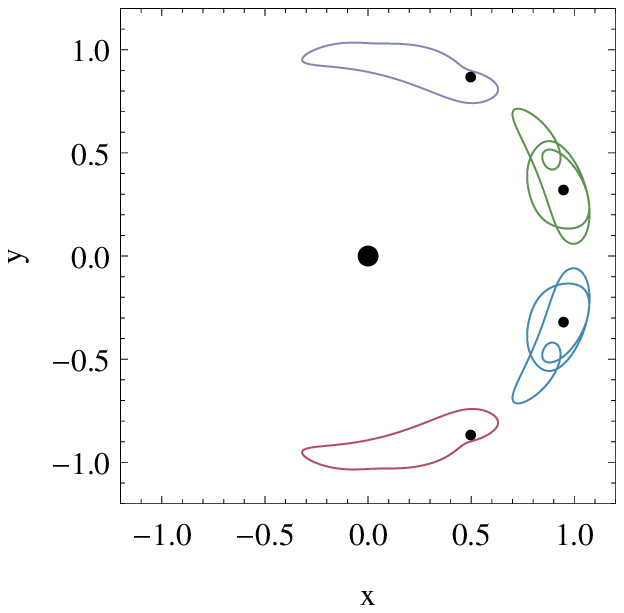}}
\caption{\label{lastpage}\label{fig:c1}The family $c1$}
\end{figure}

This section provides example orbits for the families associated with the $n=4$ equilibria. The primary families $\mathrm{\alpha Ia_4}$, $\mathrm{\beta Ia_4}$, $\mathrm{\gamma Ia_4}$, $\mathrm{\alpha II_4}$, $\mathrm{\alpha IIIa_4}$ and $\mathrm{\beta IIIa_4}$ are shown in Figs. \ref{fig:a1} to \ref{fig:a3}. The secondary connecting families are shown in Figs. \ref{fig:b1} to \ref{fig:c1}. As discussed in the text these families are shown for high mass ratio of $10^{-2}$ in order to exaggerate the radial extent of the orbit.

\begin{figure}
\centering
\subfigure[$\sqrt m P =0.986$]{\includegraphics[width=0.48\columnwidth]{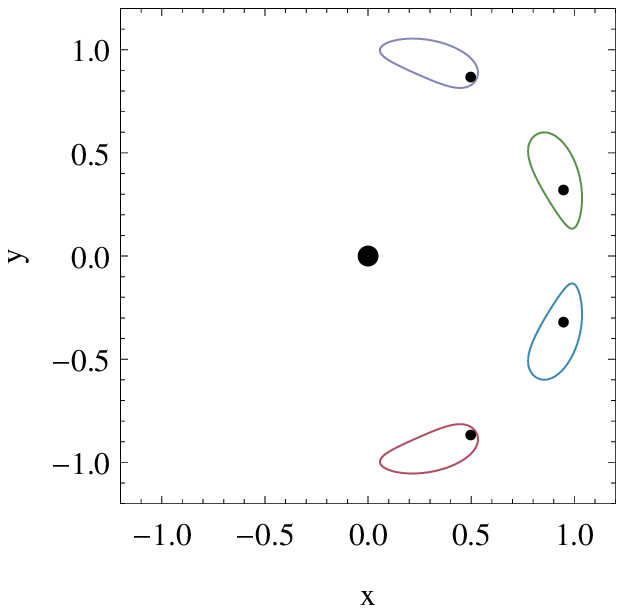}}
\subfigure[$\sqrt m P =1.769$]{\includegraphics[width=0.48\columnwidth]{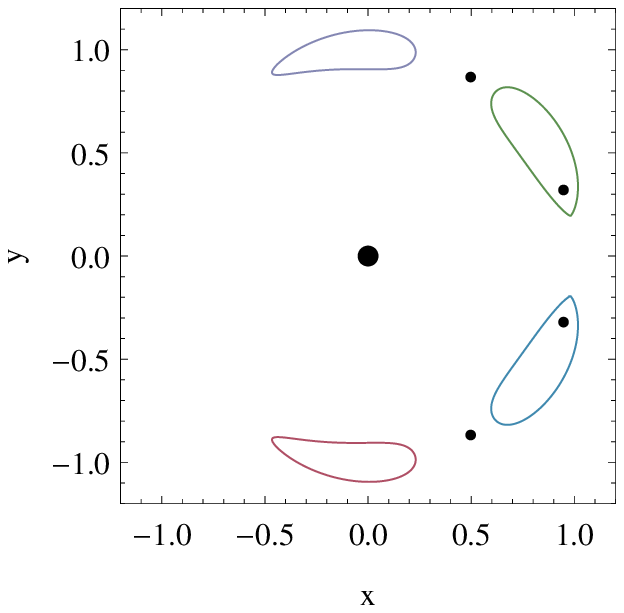}}
\subfigure[$\sqrt m P =3.123$]{\includegraphics[width=0.48\columnwidth]{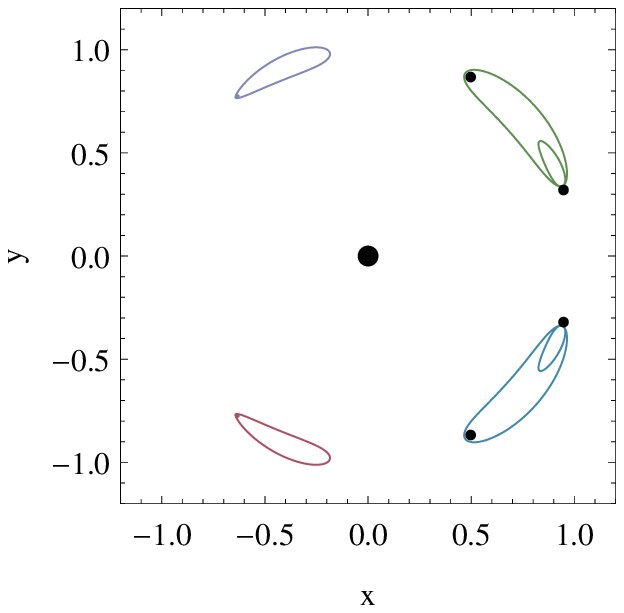}}
\subfigure[$\sqrt m P =4.606$]{\includegraphics[width=0.48\columnwidth]{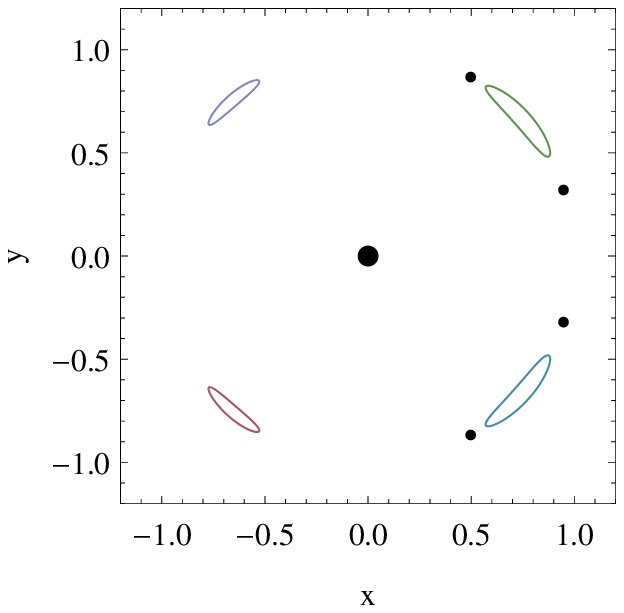}}
\caption{The family $\mathrm{\beta Ia_4}$}
\end{figure}

\begin{figure}
\centering
\subfigure[$\sqrt m P =1.698$]{\includegraphics[width=0.48\columnwidth]{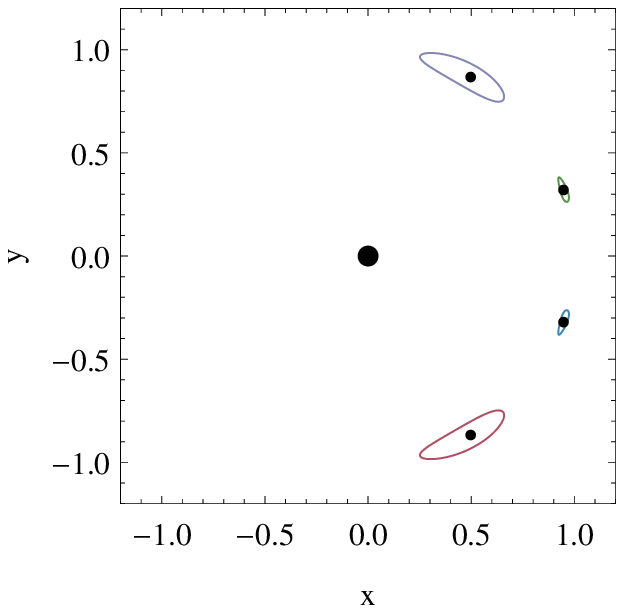}}
\subfigure[$\sqrt m P =2.513$]{\includegraphics[width=0.48\columnwidth]{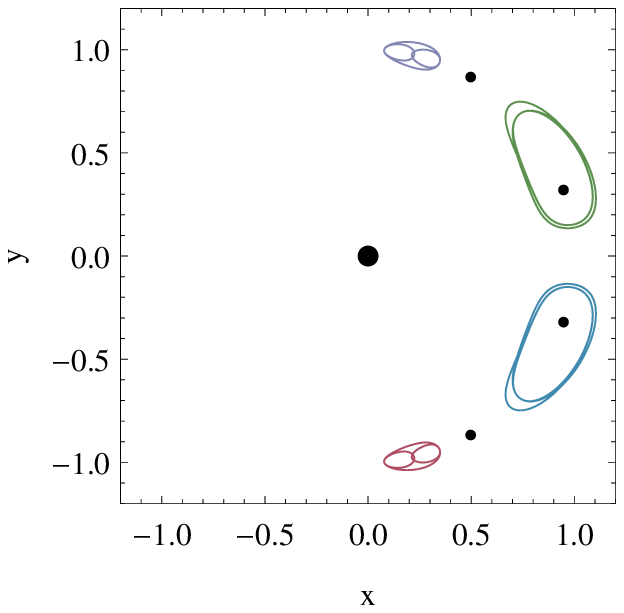}}
\subfigure[$\sqrt m P =3.736$]{\includegraphics[width=0.48\columnwidth]{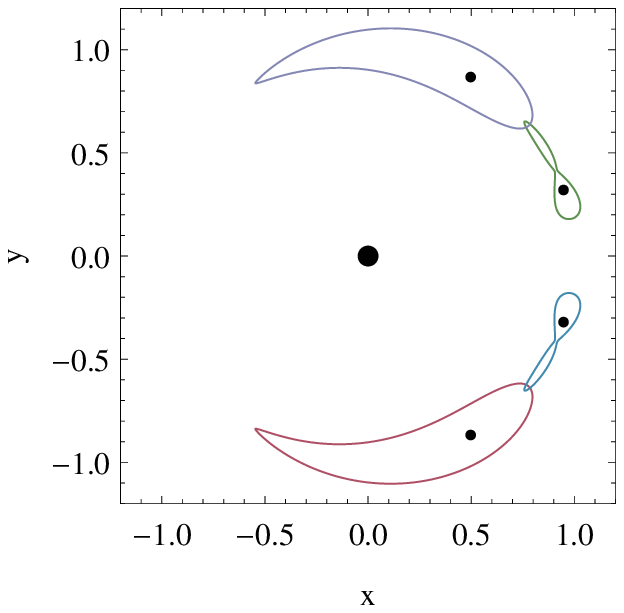}}
\subfigure[$\sqrt m P =5.022$]{\includegraphics[width=0.48\columnwidth]{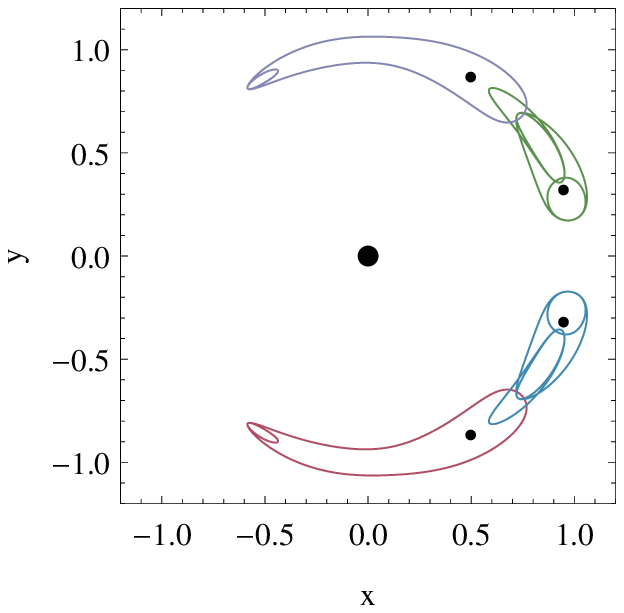}}
\subfigure[$\sqrt m P =3.704$]{\includegraphics[width=0.48\columnwidth]{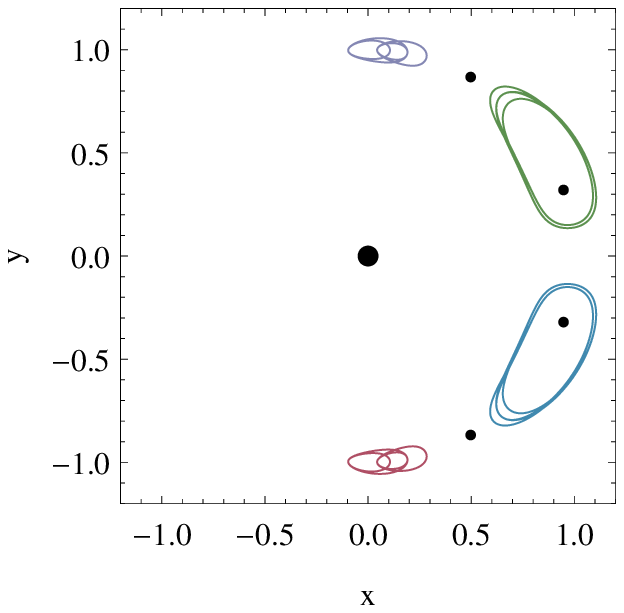}}
\subfigure[$\sqrt m P =3.681$]{\includegraphics[width=0.48\columnwidth]{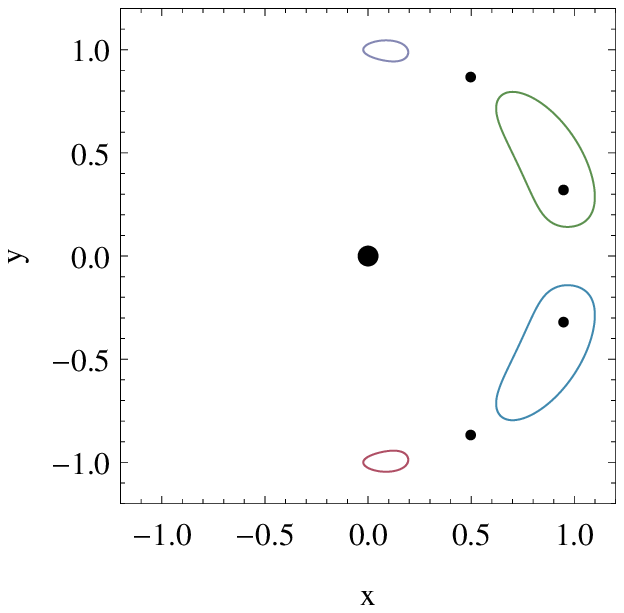}}
\caption{The family $\mathrm{\gamma Ia_4}$}
\end{figure}

\begin{figure}
\centering
\subfigure[$\sqrt m P =1.640$]{\includegraphics[width=0.48\columnwidth]{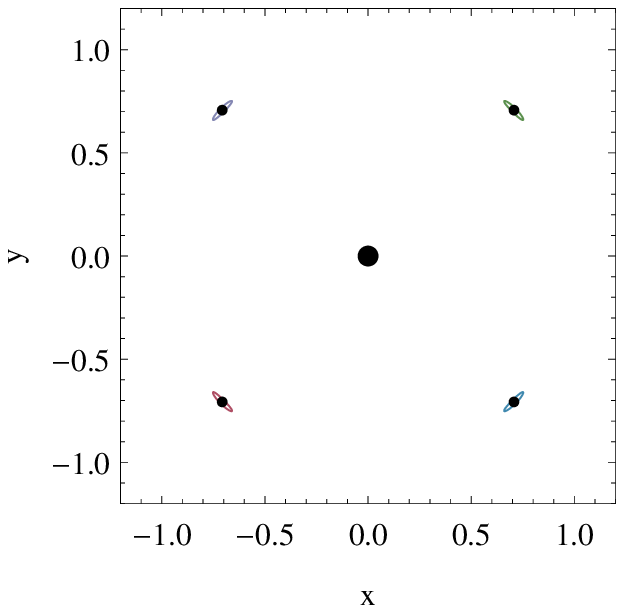}}
\subfigure[$\sqrt m P =3.080$]{\includegraphics[width=0.48\columnwidth]{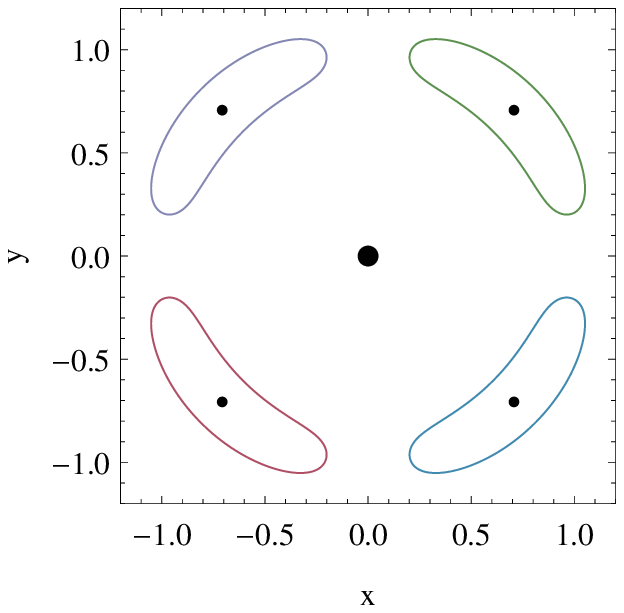}}
\caption{The family $\mathrm{\alpha II_4}$}
\end{figure}

\begin{figure}
\centering
\subfigure[$\sqrt m P =1.509$]{\includegraphics[width=0.48\columnwidth]{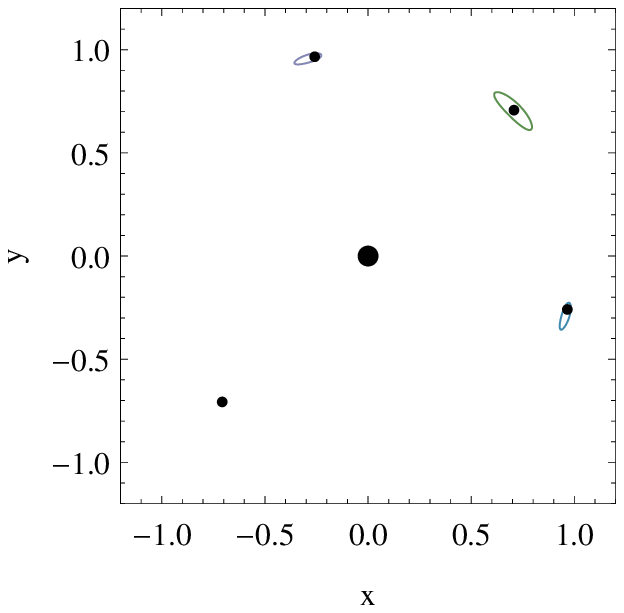}}
\subfigure[$\sqrt m P =1.683$]{\includegraphics[width=0.48\columnwidth]{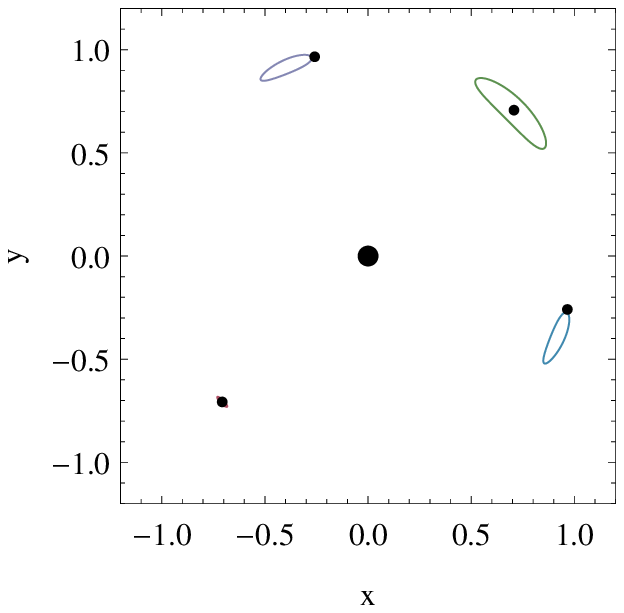}}
\subfigure[$\sqrt m P =1.952$]{\includegraphics[width=0.48\columnwidth]{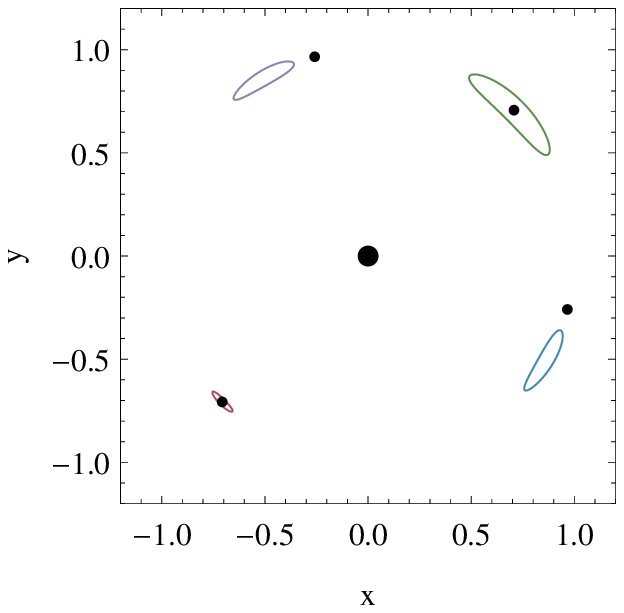}}
\subfigure[$\sqrt m P =2.385$]{\includegraphics[width=0.48\columnwidth]{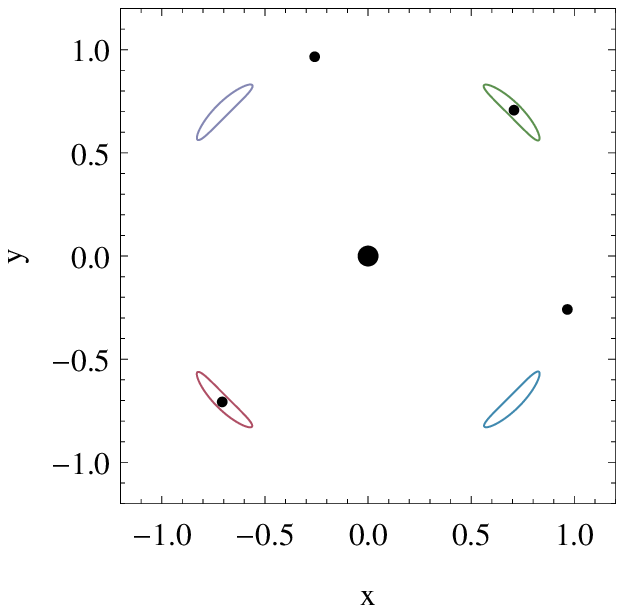}}
\caption{The family $\mathrm{\alpha IIIa_4}$}
\end{figure}

\begin{figure}
\centering
\subfigure[$\sqrt m P =2.959$]{\includegraphics[width=0.48\columnwidth]{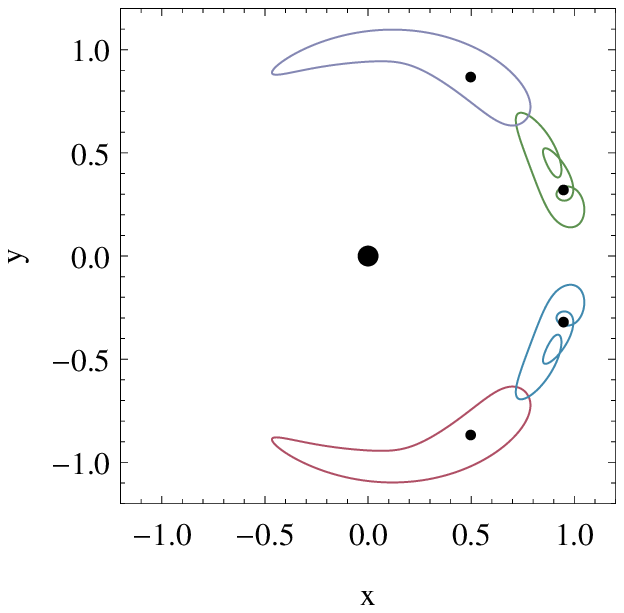}}
\subfigure[$\sqrt m P =2.489$]{\includegraphics[width=0.48\columnwidth]{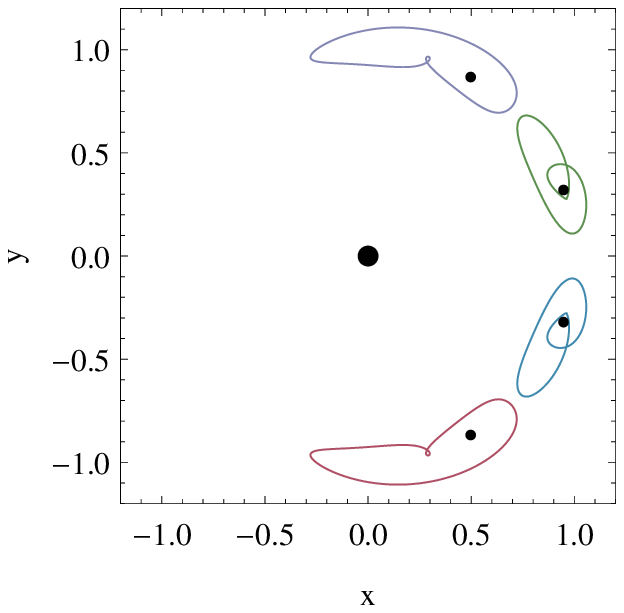}}
\subfigure[$\sqrt m P =2.233$]{\includegraphics[width=0.48\columnwidth]{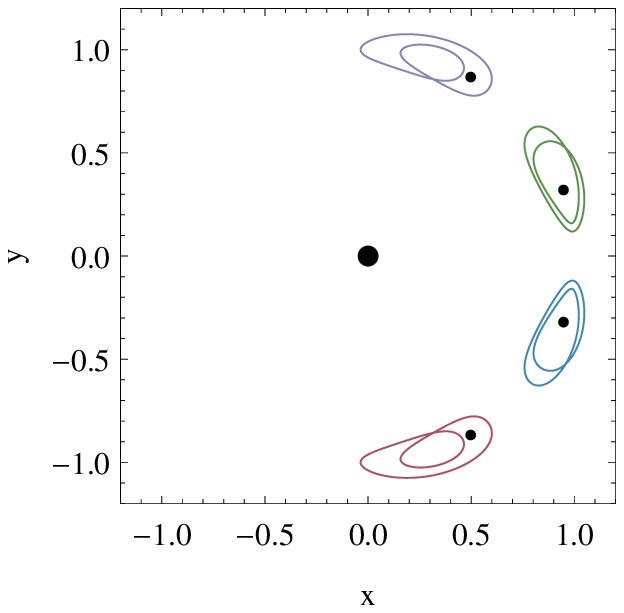}}
\subfigure[$\sqrt m P =2.206$]{\includegraphics[width=0.48\columnwidth]{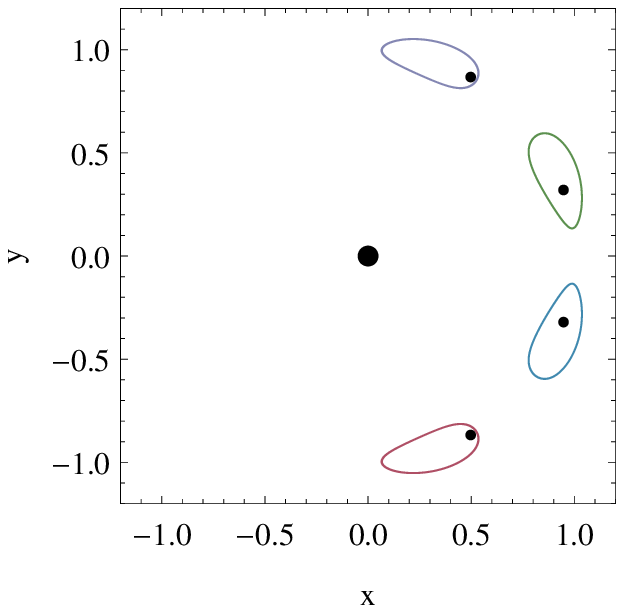}}
\caption{The family $\gamma 1$}
\end{figure}

\begin{figure}
\centering
\subfigure[$\sqrt m P =2.608$]{\includegraphics[width=0.48\columnwidth]{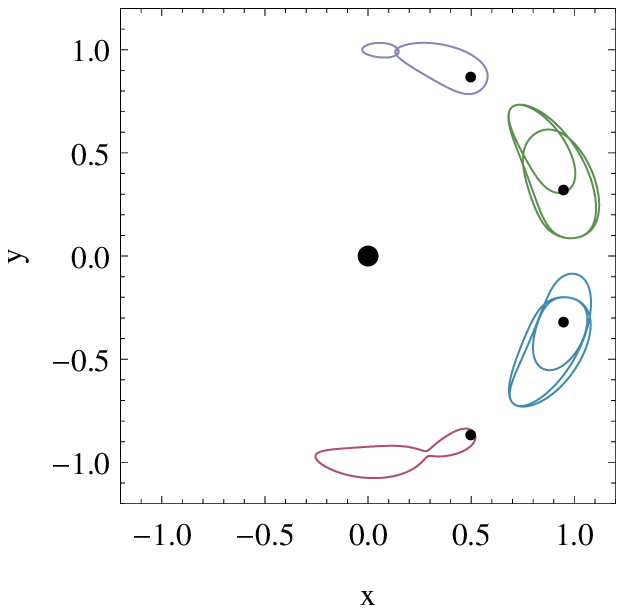}}
\subfigure[$\sqrt m P =2.844$]{\includegraphics[width=0.48\columnwidth]{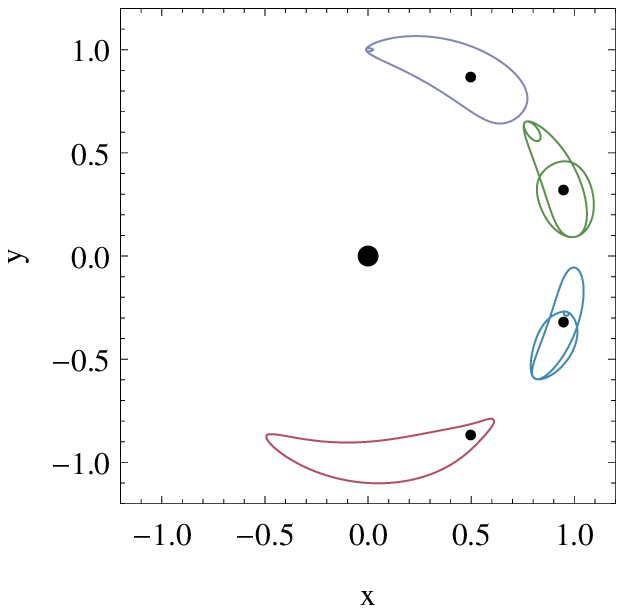}}
\subfigure[$\sqrt m P =4.047$]{\includegraphics[width=0.48\columnwidth]{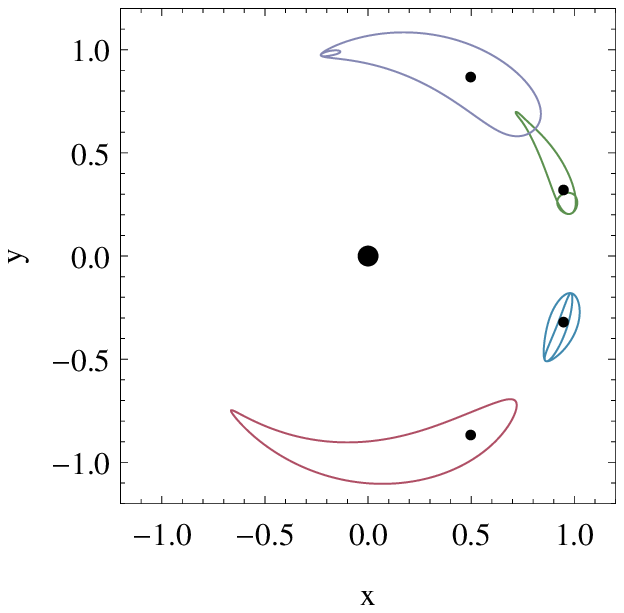}}
\subfigure[$\sqrt m P =5.320$]{\includegraphics[width=0.48\columnwidth]{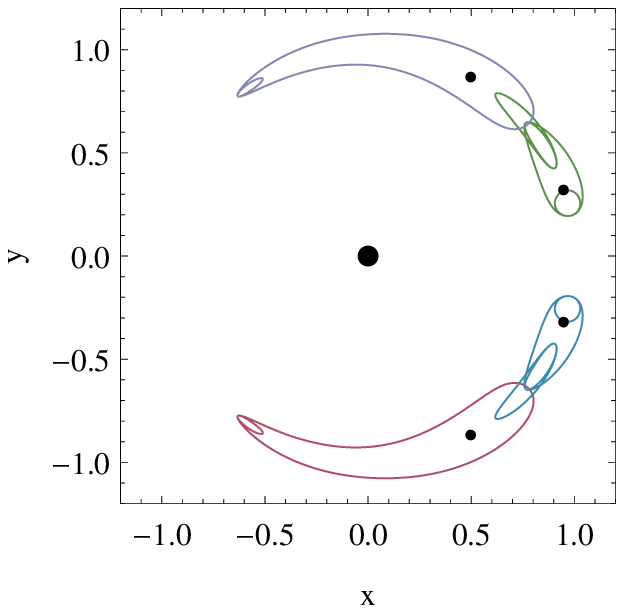}}
\caption{The family $\gamma 2$}
\end{figure}

\begin{figure}
\centering
\subfigure[$\sqrt m P =2.643$]{\includegraphics[width=0.48\columnwidth]{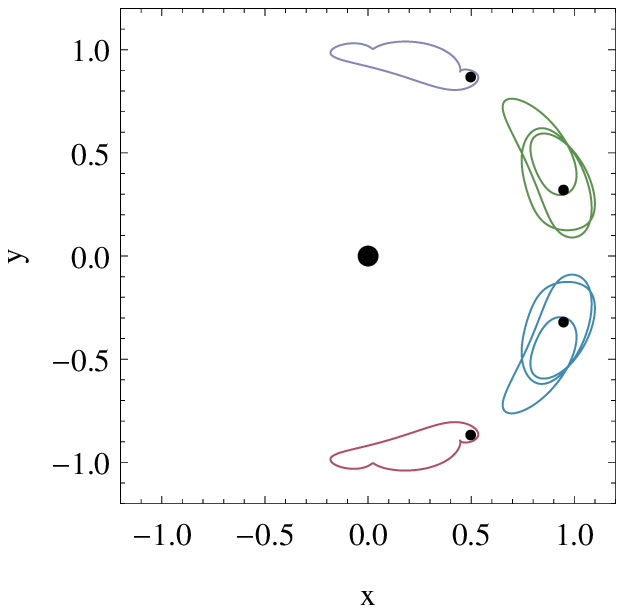}}
\subfigure[$\sqrt m P =3.355$]{\includegraphics[width=0.48\columnwidth]{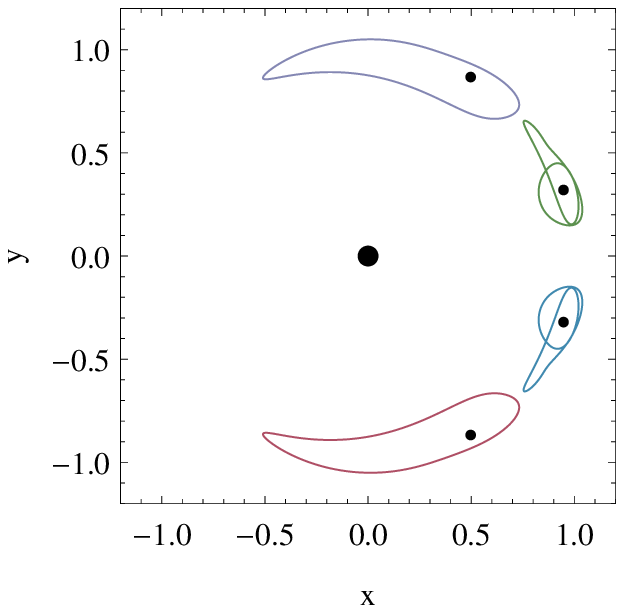}}
\subfigure[$\sqrt m P =5.271$]{\includegraphics[width=0.48\columnwidth]{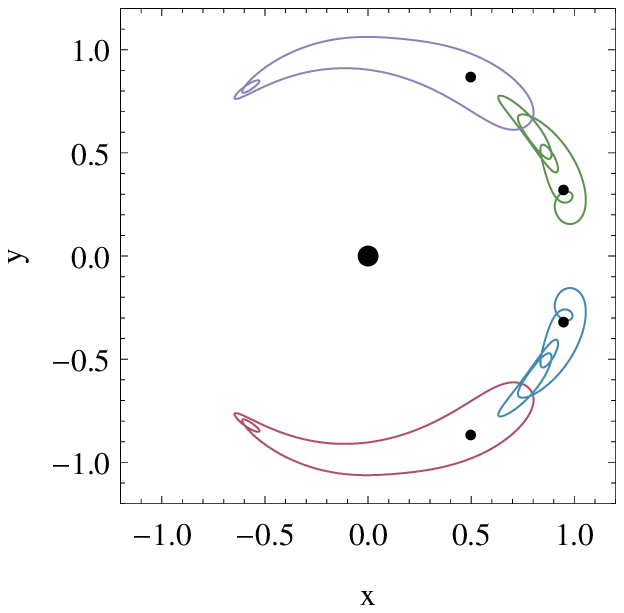}}
\subfigure[$\sqrt m P =5.376$]{\includegraphics[width=0.48\columnwidth]{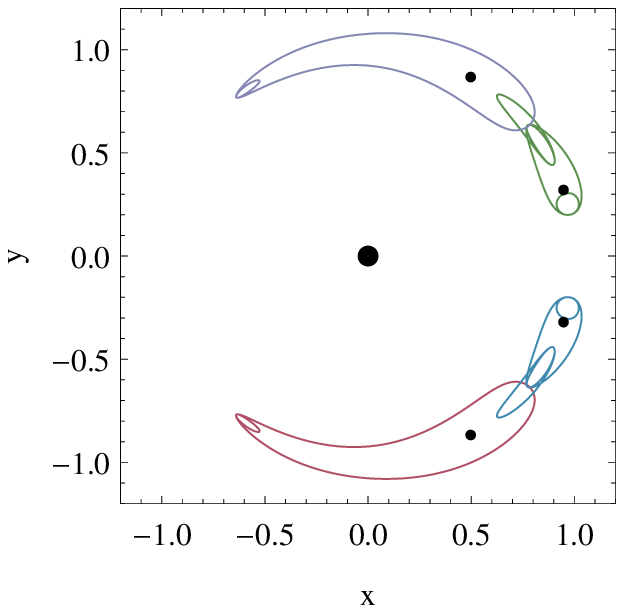}}
\caption{The family $\gamma 3$}
\end{figure}

\begin{figure}
\centering
\subfigure[$\sqrt m P =3.038$]{\includegraphics[width=0.48\columnwidth]{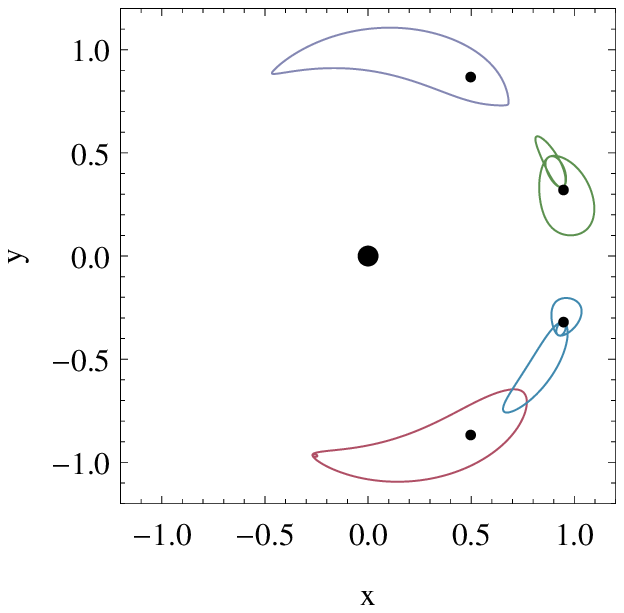}}
\subfigure[$\sqrt m P =2.659$]{\includegraphics[width=0.48\columnwidth]{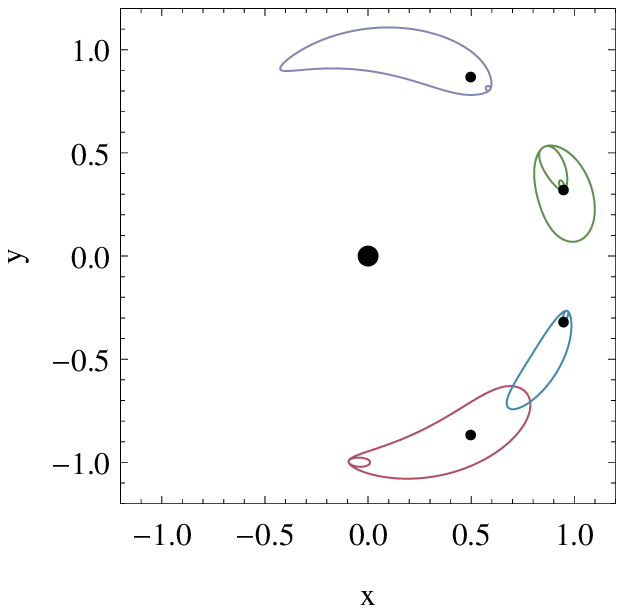}}
\subfigure[$\sqrt m P =2.354$]{\includegraphics[width=0.48\columnwidth]{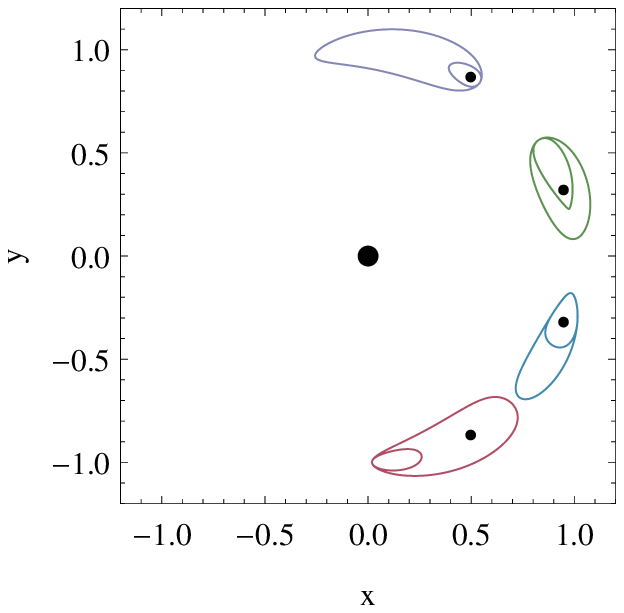}}
\subfigure[$\sqrt m P =2.206$]{\includegraphics[width=0.48\columnwidth]{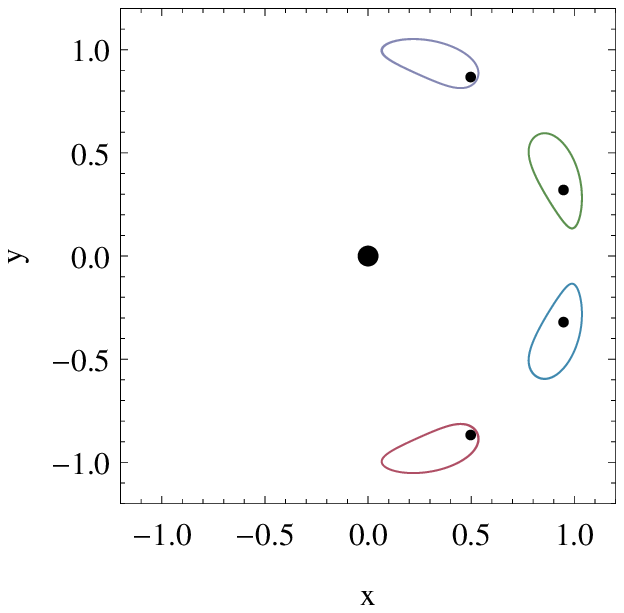}}
\caption{The family $\gamma 4$}
\end{figure}

\bsp



\begin{thebibliography}{99}

\bibitem[Albouy et al.(2012)]{Albouy2012}
Albouy, A., Cabral, H.~E., Santos, A.~A.: Some problems on the classical n-body problem, Celestial Mechanics and Dynamical Astronomy, 113, 369 (2012)

\bibitem[Albouy \& Fu(2009)]{Albouy2009}
Albouy, A., Fu, Y.: Relative equilibria of four identical satellites, Royal Society of London Proceedings Series A, 465, 2633 (2009)

\bibitem[Beaug{\'e} et al.(2007)]{Beauge2007} 
Beaug{\'e}, C., S{\'a}ndor, Z., {\'E}rdi, B., S\"uli, {\'A}.\ 2007, A\& A, 463, 359 

\bibitem[Calleja et al.(2012)]{Calleja2012}
Calleja, R.~C., Doedel, E.~J., Humphries, A.~R., Lemus-Rodr{\'{\i}}guez, A., Oldeman, E.~B.: Boundary-value problem formulations for computing invariant manifolds and connecting orbits in the circular restricted three body problem, Celestial Mechanics and Dynamical Astronomy, 114, 77 (2012)

\bibitem[Casasayas et al.(1994)]{Casasayas1994}
Casasayas, Llibre \& Nunes \ 1994, \textit{Central configurations of the planar 1+N body problem}, CeMDA 60, 274

\bibitem[Corbera et al.(2011)]{Cors2011}
Corbera, Cors \& Llibre \ 2011, \textit{On the central configurations of the planar 1+3 body problem}, CeMDA 109, 27 

\bibitem[Cors, Llibre \& Oll\'e(2004)]{Cors2004}
Cors, Llibre \& Oll\'e \ 2004, \textit{Central configurations of the planar coorbital satellite problem},  CeMDA, 89, 319

\bibitem[Doedel et al.(2011)]{Doedel2011} 
Doedel, E.~J. et al.: AUTO-07p: continuation and bifurcation software for ordinary differential equations (2011)

\bibitem[Doedel et al.(2007)]{Doedel2007} 
Doedel, E.~J., Romanov, V.~A., Paffenroth, R.~C., Keller, H.~B., Dichmann, D.~J. , Gal\'{a}n-Vioque, J. ,Vanderbauwhede, A. : Elemental periodic orbits associated with the libration points in the circular restricted 3-body problem, Int. J. Bifurc. Chaos, 17, 2625--2678 (2007)

\bibitem[Doedel et al.(2005)]{Doedel2005} 
Doedel, E.~J., Govaerts, W., Kuznetsov, Y.~A., Dhooge, A.: Numerical continuation of branch points of equilibria and periodic orbits, Int. J. Bifurc. Chaos, 15, 841--860 (2005) 

\bibitem[Doedel et al.(2003)]{Doedel2003}
Doedel, E.~J., Paffenroth, R.~C., Keller, H.~B., Dichmann, D.~J., Gal\'{a}n-Vioque, J., Vanderbauwhede, A.: Computation of periodic solutions of conservative systems with application to the 3-body problem, Int. J. Bifurc. Chaos, 13, 1353 (2003)

\bibitem[Doedel et al.(1991)]{Doedel1991}
Doedel, E.~J., Keller, H.~B., Kern\'{e}vez, J.~P.: Numerical analysis and control of bifurcation problems (I): bifurcation in finite dimensions, Int. J. Bifurc. Chaos, 1, 493--520 (1991)

\bibitem[Garc\'ia Y\'arnoz et al.(2013)]{Yarnoz2013}
Garc\'ia Y\'arnoz, D., S\'anchez, J.~P., McInnes, C.: Easily retrievable objects among the NEO population, Celestial Mechanics and Dynamical Astronomy, 116, 367 (2013)

\bibitem[Giuppone et al.(2012)]{Giuppone2012} 
Giuppone, C.~A., Ben{\'{\i}}tez-Llambay, P., \& Beaug{\'e}, C.\ 2012, MNRAS, 421, 356 

\bibitem[Laughlin \& Chambers(2002)]{laughlin} 
Laughlin, G., \& Chambers, J.~E.\ 2002, AJ, 124, 592 

\bibitem[Madhusudhan \& Winn(2009)]{Madhusudhan2009} 
Madhusudhan, N., \& Winn, J.~N.\ 2009, ApJ, 693, 784 

\bibitem[Maxwell(1890)]{Maxwell1890} 
Maxwell, C.~J.: On the stability of the motion of Saturn's Rings, in Scientific Papers of James Clerk Maxwell, Cambridge University Press

\bibitem[Moeckel(1994)]{Moeckel1994}
Moeckel \ 1994, \noindent\textit{Linear stability of relative equilibria with a dominant mass}, J. Dynamics \& Diff. Eq., 6, 37

\bibitem[Mun\~{o}z-Almaraz et al.(2007)]{MunozAlmaraz2007}
Mu{\~n}oz-Almaraz, F.~J., Freire, E., Galan-Vioque, J., Vanderbauwhede, A.: Continuation of normal doubly symmetric orbits in conservative reversible systems, Celestial Mechanics and Dynamical Astronomy, 97, 17 (2007)

\bibitem[Mun\~{o}z-Almaraz et al.(2003)]{Munoz2003} 
Mun\~{o}z-Almaraz, F.~J., Freire, E., Gal{\'a}n, J., Doedel, E., Vanderbauwhede, A.: Continuation of periodic orbits in conservative and Hamiltonian systems, Phys. D Nonlinear Phenom., 181, 1--38 (2003)

\bibitem[Murray \& Dermott(1999)]{MD1999}
Murray, C.~D., Dermott, S.~F.: Solar system dynamics, Cambridge University Press, Cambridge (1999)

\bibitem[Press et al.(1989)]{Press}
Press W.~H., Flannery B.~P., Teukolsky S., Vetterling W.~T.: Numerical recipes in C, Cambridge University Press, Cambridge (1989)

\bibitem[Renner \& Sicardy(2004)]{RS2004} Renner, S., \& Sicardy, B.\ 2004, \textit{Stationary Configurations for Co-orbital Satellites with Small Arbitrary Masses}, Celestial Mechanics and Dynamical Astronomy, 88, 397 

\bibitem[Salo \& Yoder(1988)]{Salo1988}
Salo \& Yoder \ 1988, \textit{The dynamics of coorbital satellite systems}, A\& A, 205, 309

\bibitem[Schwarz et al.(2009)]{Schwarz2009} 
Schwarz, R., S{\"u}li, {\'A}., Dvorak, R., \& Pilat-Lohinger, E.\ 2009, Celestial Mechanics and Dynamical Astronomy, 104, 69 

\bibitem[Schwarz et al.(2007)]{Schwarz2007} 
Schwarz, R., Dvorak, R., S{\"u}li, {\'A}., \& {\'E}rdi, B.\ 2007, A\& A, 474, 1023 

\bibitem[Yoder et al.(1983)]{Yoder1983}
Yoder, C.~F., Colombo, G., Synnott, S.~P., Yoder, K.~A.: Theory of motion of Saturn's coorbiting satellites, Icarus, 53, 431 (1983)

\end{thebibliography}
\end{document}